\documentclass[preprint,aps]{revtex4}
\usepackage{graphicx}
\usepackage{dcolumn}
\usepackage{bm}
\usepackage{amssymb}
\usepackage{amsmath}
\newtheorem{theorem}{Theorem}
\newtheorem{corollary}{Corollary}
\newtheorem{lemma}{Lemma}
 \newenvironment{Proof.}[1][Proof.]{\begin{trivlist}
     \item[\hskip \labelsep {\bfseries #1}]}{\end{trivlist}}
\begin{document}
\setcounter{page}{1}
\title
{On the existence of certain axisymmetric interior metrics}
\author
{C. Angulo Santacruz$^{1}$, D. Batic$^{1,}$$^2$ and M.
Nowakowski$^3$} \affiliation{ $^1$ Departamento de Matematicas,
Universidad de los
Andes, Cra.1E No.18A-10, Bogota, Colombia\\
$^2$ Department of Mathematics, University of West
Indies, Kingston 6, Jamaica\\
$^3$ Departamento de Fisica, Universidad de los Andes, Cra.1E
No.18A-10, Bogota, Colombia}

\begin{abstract}
One of the effects of noncommutative coordinate operators is that
the delta-function connected to the quantum mechanical amplitude
between states sharp to the position operator gets smeared by a
Gaussian distribution. Although this is not the full account of
effects of noncommutativity, this effect is in particular
important, as it removes the point singularities of Schwarzschild
and Reissner-Nordstr\"{o}m solutions. In this context, it seems to
be of some importance to probe also into ring-like singularities
which appear in the Kerr case. In particular, starting with an
anisotropic energy-momentum tensor and a general axisymmetric
ansatz of the metric together with an arbitrary mass distribution
(e.g. Gaussian) we derive the full set of Einstein equations that
the Noncommutative Geometry inspired Kerr solution should satisfy.
Using these equations we prove two theorems regarding the
existence of certain Kerr metrics inspired by Noncommutative
Geometry.
\end{abstract}

\maketitle

\section{Introduction}
Noncommutative Geometry \cite{Rest1} has been an active field of
research in the last decades. In particular, there have been
several attempts \cite{three,four} to cure the appearance of
infinities in Quantum Field Theory by means of noncommutative
field theories where the formulation is done in terms of ordinary
functions of commuting variables endowed with a Moyal or
$*$-product. However, in the formulation of Noncommutative
Geometry based on the aforementioned product the basic property of
the noncommutativity, namely the existence of a natural
ultraviolet cutoff due to the uncertainty in position, is not
transparent. In particular, the free propagator is unaffected by
the $*$-product as if noncommutativity had no effect on it.
Calculations are than performed using truncated series expansion
in the parameter characterizing the noncommutativity of space.
This approach leads to the ultraviolet/infrared mixing phenomenon
but one is still faced with divergences \cite{peet} to be cured as
in ordinary Quantum Field Theory. At the present stage, we cannot
exclude that the complete summation of the $*$-product expansion
might give a finite result but such a procedure presents two major
difficulties: it is far from being easy and goes beyond
calculational capabilities. Therefore, even if the use of the
Moyal product is a well defined procedure, its application as for
now does not produce ultraviolet finiteness of Noncommutative
Field Theory. On the other hand, a new formulation of Quantum
Field Theory on the noncommutative plane \cite{spallucci1}
indicated that the existence of a minimal length in a
noncommutative plane manifests itself already in the free
propagator. The core ingredients in the formulation of this model
without $*$-product is the use of expectation values of operators
between \textit{coherent states} \cite{Glauber}, whereas the
noncommutativity of space-time is encoded in the commutator $
[{\bf{x}}^{\mu},{\bf{x}}^{\nu}]=i\Theta^{\mu\nu}$, where
$\Theta^{\mu\nu}$ is an anti-symmetric matrix which determines the
fundamental cell discretization of space-time. We recall that a
similar commutation relation has been already introduced in the
seminal paper of Snyder \cite{snyder}. In this approach,
noncommutativity of coordinates is carried on by the Gaussian
spread of coherent states and modifies the structure of the
Feynman propagator rendering Quantum Field Theory ultraviolet
finite \cite{spallucci1}. To be more specific, the amplitude
$\langle \mathbf{z} \vert \mathbf{x} \rangle$ which in the
standard quantum mechanical prescription is proportional to the
Dirac delta function $\delta^{(3)}(\mathbf{z} - \mathbf{x})$ can
be replaced in Noncommutative Geometry by a Gaussian distribution
proportional to $exp\left(-(\mathbf{z}
-\mathbf{x})^2/2\theta\right)$, where $\theta$ is a tiny parameter
encoding the noncommutativity of the space-time fabric
\cite{spallucci11}. Here, it must be emphasized that such a
replacement is also possible when one considers the Voros product
instead of the Moyal product. For such a discussion, we refer to
\cite{indian}.\\
One might suspect that noncommutativity might also cure
divergences appearing in various forms in General Relativity. This
conjecture has been studied in
\cite{indian,nico1,nico2,spallucci3,ansoldi1,nico3,uns}, where the
noncommutative inspired counterparts of the Schwarzschild and
Reissner-Nordstr\"{o}m metrics have been derived. All these new
black hole geometries contain their respective classical solutions
in the limit of a vanishing noncommutativity parameter $\theta$.
Moreover, the central singularity is now replaced by a regular
region represented by a self-gravitating, droplet of anisotropic
fluid. The main idea behind the derivation of these solutions does
not rely on a modification of the four-dimensional Einstein action
to incorporate noncommutative effects but it consists in
implementing noncommutativity only in the matter source
\cite{nico3}. In the present work we specialize to the
axisymmetric case of the Kerr black hole where one would expect
that the classical ring singularity should be replaced by a
certain Gaussian distribution given in appendix D. In particular,
we present the full set of Einstein equations with anisotropic
energy-momentum tensor. The resulting set of equations is a highly
complicated system of nonlinear partial differential equations,
for which at present we do not have a solution. We recall however,
that this is also the case for a similar set of equations where in
contrast to us the energy-momentum tensor is isotropic
\cite{chandra1}. The usefulness of our equations lies in the fact
that any noncommutative Kerr candidate, be it an educated guess or
a particular solution, must satisfy the set of equations
(\ref{G1})-(\ref{hyd2}). In this respect, we mention the work of
\cite{spallucci2} where the authors claimed to have derived the
noncommutative counterpart of the Kerr metric. Despite the
educated guess leading to the so-called Kerrr metric the authors
did not perform any consistency check regarding the question
whether their solution satisfies \textit{all} Einstein field
equations coupled to an anisotropic source or not. Here, we offer
the solution to this problem presented in form of two theorems. In
appendix B we performed the consistency check for the same choice
of the energy-momentum tensor done in \cite{spallucci2} and we
showed that the metric represented by equation (44) in
\cite{spallucci2} can never satisfy all Einstein field equations.

\section{The Einstein field equations with anisotropic matter source}
According to \cite{chandra} the most general ansatz for metrics
describing stationary axisymmetric space-times in Boyer-Lindquist
coordinates $(t,R,\vartheta,\varphi)$ can be written in the form
\begin{equation}\label{sam}
ds^2=(e^{2\nu}-\omega^2e^{2\psi})dt^2+2\omega e^{2\psi}dtd\varphi
-e^{2\mu_R}dR^2-e^{2\mu_\vartheta}d\vartheta-e^{2\psi}d\varphi^2
\end{equation}
where $\nu,\psi,\mu_R$ and $\mu_\vartheta$ are in general
functions of the spheroidal coordinates $R$ and $\vartheta$
defined through the relations \cite{Boyer}
\[
\frac{x^2+y^2}{R^2+a^2}+\frac{z^2}{R^2}=1,\quad
\frac{x^2+y^2}{a^2\sin^2{\vartheta}}-\frac{z^2}{a^2\cos^{2}{\vartheta}}=1,
\]
where $a$ is a real parameter. Furthermore, we consider the
Einstein field equations
\begin{equation}\label{EinsteinFieldEquations}
G_{\mu\nu}=-8\pi T_{\mu\nu},
\end{equation}
where the energy-momentum tensor $T_{\mu\nu}$ describes an
anisotropic perfect fluid with time-like four velocity $u_{\mu}$,
mass density $\rho$ and pressures $p_R$, $p_\vartheta$. We shall
derive the equations governing any stationary, axial-symmetric
solution of Einstein equations (\ref{EinsteinFieldEquations}) in
the presence of an anisotropic perfect fluid, where the mass
density $\rho$ is assigned from the very beginning in the spirit
of \cite{nico1,nico2,spallucci3,ansoldi1,nico3,spallucci2}. More
precisely, $T_{\mu\nu}$ reads ($c=G=1$ units)
\begin{equation}\label{enmomten}
T_{\mu\nu}=(\rho+p_\vartheta)u_{\mu}u_{\nu}-p_\vartheta
g_{\mu\nu}+(p_R-p_\vartheta)\ell_\mu \ell_\nu
\end{equation}
where $\ell^\mu$ is a unit space-like vector orthogonal to the
fluid four-velocity $u^\mu$, i.e.
\[
\ell^\mu \ell_\mu=-1,\quad \ell^\mu u_{\mu}=0.
\]
Moreover, the fluid four-velocity has to satisfy the condition
\begin{equation}\label{g_m}
g_{\mu\nu}u^\mu u^\nu=1.
\end{equation}
At this point a couple of remarks are in order. First of all,
notice that our choice of $T_{\mu\nu}$ contains as a special case
the energy-momentum tensor given by equation (37) in
\cite{spallucci2}, since it reduces to (37) by taking $p_R=-\rho$.
However, in an axial-symmetric space-time there is no a priori
reason to believe that the fluid should satisfy an equation of
state of the de Sitter form as it is indeed the case for the
spherically symmetric noncommutative geometry inspired
Schwarzschild and Reissner-Nordstr\"om metrics. Furthermore, the
choice of the fluid velocities (equation (38) in
\cite{spallucci2}) does not satisfy condition (\ref{g_m}) with
$g_{\mu\nu}$ given by the Kerrr line element (44) in
\cite{spallucci2}. In fact, a simple computation shows that
\[
g_{\mu\nu}u^\mu u^\nu=(u^t)^2\left(g_{tt}+2\omega
g_{t\varphi}+\omega^2
g_{\varphi\varphi}\right)=\frac{\Delta^2}{(R^2+a^2)^2}\neq 1,
\]
where
\[
u^t=\sqrt{\frac{\Delta}{\Sigma}},\quad\omega(R)=\frac{a}{R^2+a^2},\quad\Sigma=R^2+a^2\cos^2{\vartheta}
\]
and $\Delta$ given by equation (45) in \cite{spallucci2}.
According to (\ref{g_m}), the right choice of $u^t$ in
\cite{spallucci2} should read
\[
u^t=\frac{R^2+a^2}{\sqrt{\Sigma\Delta}},
\]
but even with this choice we could prove in appendix B that the
so-called Kerrr
metric does not satisfy all Einstein field equations.\\
In the present case we specialize the energy-momentum tensor to a
stationary matter distribution with cylindrical symmetry. Hence,
there exist two Killing vector fields $\partial_t$ and
$\partial_\varphi$ and in general, there will be at least two
non-vanishing velocity components $u^t$ and $u^\varphi$, so that
the four velocity $u^\mu$ of the fluid can be written as
\cite{Ple}
\[
u^\mu=u^t(\delta^\mu_t+\Omega~\delta^\mu_\varphi),\quad
\Omega=\frac{u^\varphi}{u^t}.
\]
Here, $\Omega$ denotes the angular velocity of the fluid and
depends in general on the spheroidal variables $R$ and
$\vartheta$. Notice that the requirement $u^R=u^\vartheta=0$
ensures that the fluid is circulating within the $(t,\varphi)$ $-$
surface. Condition (\ref{g_m}) permits to express the $u^t$
component of the fluid velocity as
\[
u^t=\frac{1}{\sqrt{g_{tt}+2\Omega g_{t\varphi}+\Omega^2
g_{\varphi\varphi}}}=V^{-1},\quad
V=V(R,\vartheta)=\sqrt{e^{2\nu}-(\omega-\Omega)^2e^{2\psi}}
\]
whereas the non-vanishing covariant components of the four
velocity are
\[
u_t=V^{-1}\left[e^{2\nu}-\omega(\omega-\Omega)e^{2\psi}\right],\quad
u_\varphi=V^{-1}(\omega-\Omega)e^{2\psi}.
\]
Concerning the choice of the unit space-like vector $\ell^{\mu}$,
we observe that it will be orthogonal to the fluid four velocity
if we take $\ell^t=\ell^\varphi=0$. Moreover, the condition that
$\ell^\mu$ be a unit space-like vector requires that
\[
e^{2\mu_R}(\ell^R)^2+e^{2\mu_\vartheta}(\ell^\vartheta)^2=1.
\]
Notice that the above equation does not uniquely fix $\ell^\mu$
and we can follow at least two different approaches in determining
the components of $\ell^\mu$. The first approach is standard and
gives rise to the choice
\begin{equation}\label{L1}
\ell^\mu=e^{-\mu_R}~\delta^\mu_R,\quad \ell^\vartheta=0,
\end{equation}
whereas the second one is represented by the more sophisticated
expression
\begin{equation}\label{L2}
\ell^\mu=\frac{1}{\sqrt{2}}\left(e^{-\mu_R}~\delta^\mu_R+e^{-\mu_\vartheta}~\delta^\mu_\vartheta\right).
\end{equation}
In order to derive the full set of Einstein field equations
corresponding to each of the possible choices (\ref{L1}) and
(\ref{L2}), it is convenient to start with the following
equivalent form of (\ref{EinsteinFieldEquations}), namely
\begin{equation}\label{efe}
R_{\mu\nu}=8\pi\left(\frac{T}{2}~g_{\mu\nu}-T_{\mu\nu}\right),\quad
T=g^{\mu\nu}T_{\mu\nu}
\end{equation}
together with the conservation equation
\begin{equation}\label{conservazioneTmunu}
T^{\mu\nu}{}_{;\nu}=0.
\end{equation}
However, in the coordinate approach the components of the Ricci
tensor assume a very complicated form. For this reason, we shall
follow the line of reasoning in \cite{chandra1,chandra} and
introduce the tetrad-frame
\[
e_{(t)}=e^{-\nu}\partial_t+\omega e^{-\nu}\partial_{\varphi},\quad
e_{(\varphi)}=e^{-\psi}\partial_\varphi,\quad
e_{(R)}=e^{-\mu_R}\partial_R,\quad
e_{(\vartheta)}=e^{-\mu_\vartheta}\partial_\vartheta.
\]
In this setting the relevant Einstein field equations for our
problem are
\begin{equation}\label{alpha1}
R_{(t)(t)}=8\pi\left(\frac{T}{2}-T_{(t)(t)}\right),\quad
R_{(t)(\varphi)}=-8\pi T_{(t)(\varphi)},
\end{equation}
\begin{equation}\label{alpha2}
R_{(\varphi)(\varphi)}=-8\pi\left(\frac{T}{2}+T_{(\varphi)(\varphi)}\right),\quad
G_{(R)(R)}=-8\pi T_{(R)(R)},
\end{equation}
\begin{equation}\label{alpha3}
R_{(R)(\vartheta)}=-8\pi T_{(R)(\vartheta)},\quad
G_{(\vartheta)(\vartheta)}=-8\pi T_{(\vartheta)(\vartheta)}.
\end{equation}
Notice that the vanishing of the component $T_{(R)(\vartheta)}$ of
the energy-momentum tensor depends on the particular choice of the
unit space-like vector $\ell^\mu$, whereas the trace of
$T^{\mu\nu}$ is for both choices of $\ell^\mu$ given by
\[
T=g_{\mu\nu}T^{\mu\nu}=\rho-p_R-2p_{\vartheta}.
\]

\section{Einstein field equations with $\ell^t=\ell^\varphi=\ell^\vartheta=0$}
The non-vanishing contravariant components of the energy-momentum
tensor in the coordinate representation are
\[
T^{tt}=V^{-2}\left[\rho+(\omega-\Omega)^2
e^{2\psi-2\nu}p_{\vartheta}\right],
\]
\[
T^{t\varphi}=V^{-2}\{\
\Omega\rho-(\omega-\Omega)\left[1-\omega(\omega-\Omega)e^{2\psi-2\nu}\right]p_\vartheta\}\
,
\]
\[
T^{\varphi\varphi}=V^{-2}\{\
\Omega^2\rho+\left[e^{\nu-\psi}-\omega(\omega-\Omega)e^{\psi-\nu}\right]^2p_\vartheta\}\
,\quad T^{RR}=e^{-2\mu_R}p_{R},\quad
T^{\vartheta\vartheta}=e^{-2\mu_\vartheta}p_{\vartheta}.
\]
Taking into account that the contravariant components of the
energy-momentum tensor in the non-coordinate basis can be computed
from the relation $T^{(a)(b)}=e^{(a)}{}_\mu e^{(b)}{}_\nu
T^{\mu\nu}$ with $e^{(a)}{}_\mu$ defined in \ref{vierbeincomp}, we
find
\[
T^{(t)(t)}=V^{-2}\left[e^{2\nu}\rho+(\omega-\Omega)^2
e^{2\psi}p_{\vartheta}\right],\quad
T^{(t)(\varphi)}=-V^{-2}(\omega-\Omega)e^{\psi+\nu}~(\rho+p_\vartheta),
\]
\[
T^{(\varphi)(\varphi)}=V^{-2}\left[(\omega-\Omega)^2e^{2\psi}\rho+e^{2\nu}p_{\vartheta}
\right],\quad T^{(R)(R)}=p_{R},\quad
T^{(\vartheta)(\vartheta)}=p_{\vartheta},
\]
whereas the covariant components are computed according to
$T_{(a)(b)}=\eta_{(a)(c)}\eta_{(b)(d)}T^{(c)(d)}$. Since
$\eta_{(a)(b)}=\rm{diag}(1,-1,-1,-1)$ (see \ref{vierbeincomp}), we
shall have $T_{(a)(b)}=T^{(a)(b)}$, whenever the tetrad indices
$a$ and $b$ coincide. Moreover, if $a\neq b$, the only non zero
component of the energy-momentum tensor is that with $a=t$ and
$b=\varphi$ and we get $T_{(t)(\varphi)}=-T^{(t)(\varphi)}$.
Finally, by means of (\ref{RTT})-(\ref{GTHTH}) the Einstein field
equations (\ref{alpha1})-(\ref{alpha3}) can be explicitly written
as
\[
e^{-2\mu_R}\left[\nu_{,R,R}+\nu_{,R}(\psi+\nu-\mu_R+\mu_\vartheta)_{,R}\right]
+e^{-2\mu_\vartheta}\left[\nu_{,\vartheta,\vartheta}+\nu_{,\vartheta}(\psi+\nu+\mu_R-\mu_\vartheta)_{,\vartheta}\right]
-\frac{1}{2}e^{2(\psi-\nu)}\cdot
\]
\begin{equation}\label{G1}
\left[(\omega_{,R})^2e^{-2\mu_R}+(\omega_{,\vartheta})^2e^{-2\mu_\vartheta}\right]=
-\frac{4\pi}{V^2}\left[e^{2\nu}(\rho+p_R)+(\omega-\Omega)^2
e^{2\psi}(\rho-p_R)+2e^{2\nu}p_\vartheta\right],
\end{equation}
\[
\frac{1}{2}e^{-2\psi-\mu_R-\mu_\vartheta}\left[\left(\omega_{,R}e^{3\psi-\nu-\mu_R+\mu_\vartheta}\right)_{,R}+
\left(\omega_{,\vartheta}e^{3\psi-\nu-\mu_\vartheta+\mu_R}\right)_{,\vartheta}\right]
\]
\begin{equation}\label{G2}
=-\frac{8\pi}{V^2}~(\omega-\Omega)e^{\psi+\nu}~(\rho+p_\vartheta),
\end{equation}
\[
e^{-2\mu_R}\left[\psi_{,R,R}+\psi_{,R}(\psi+\nu-\mu_R+\mu_\vartheta)_{,R}\right]+
e^{-2\mu_\vartheta}\left[\psi_{,\vartheta,\vartheta}+\psi_{,\vartheta}(\psi+\nu+\mu_R-\mu_\vartheta)_{,\vartheta}\right]
\]
\begin{equation}\label{G3}
+\frac{1}{2}e^{2(\psi-\nu)}\left[(\omega_{,R})^2e^{-2\mu_R}+(\omega_{,\vartheta})^2e^{-2\mu_\vartheta}\right]=
4\pi\left[\rho-p_R+\frac{2(\omega-\Omega)^2e^{2\psi}}{V^2}~(\rho+p_\vartheta)\right],
\end{equation}
\begin{equation}\label{G4}
(\psi+\nu)_{,R,\vartheta}-(\psi+\nu)_{,R}\mu_{R,\vartheta}-(\psi+\nu)_{,\vartheta}\mu_{\vartheta,R}+
\psi_{,R}\psi_{,\vartheta}
+\nu_{,R}\nu_{,\vartheta}=\frac{1}{2}\omega_{,R}\omega_{,\vartheta}e^{2(\psi-\nu)},
\end{equation}
\[
e^{-2\mu_R}\left[\nu_{,R}\left(\psi+\mu_\vartheta\right)_{,R}+\psi_{,R}\mu_{\vartheta,R}\right]+
e^{-2\mu_\vartheta}\left[\left(\psi+\nu\right)_{,\vartheta,\vartheta}+(\psi+\nu)_{,\vartheta}(\nu-\mu_\vartheta)_{,\vartheta}
+\psi_{,\vartheta}\psi_{,\vartheta}\right]
\]
\begin{equation}\label{G5}
+\frac{1}{4}e^{2\psi-2\nu}\left[(\omega_{,R})^2
e^{-2\mu_R}-(\omega_{,\vartheta})^2
e^{-2\mu_\vartheta}\right]=-8\pi p_{R},
\end{equation}
\[
e^{-2\mu_R}\left[\left(\psi+\nu\right)_{,R,R}+(\psi+\nu)_{,R}(\nu-\mu_R)_{,R}+\psi_{,R}\psi_{,R}\right]+
e^{-2\mu_\vartheta}\left[\nu_{,\vartheta}\left(\psi+\mu_R\right)_{,\vartheta}+\psi_{,\vartheta}\mu_{R,\vartheta}\right]
\]
\begin{equation}\label{G6}
-\frac{1}{4}e^{2\psi-2\nu}\left[(\omega_{,R})^2
e^{-2\mu_R}-(\omega_{,\vartheta})^2
e^{-2\mu_\vartheta}\right]=-8\pi p_\vartheta.
\end{equation}
Notice that the system of six equations (\ref{G1})-(\ref{G6}) is
under-determined, since it involves the eight unknowns
$\nu,\psi,\omega,\mu_R,\mu_\vartheta,p_R,p_\vartheta$ and
$\Omega$. The additional two equations needed to close the system
are provided by the conservation equation
\[
T^{\mu\nu}{}_{;\nu}
=T^{\mu\nu}{}_{,\nu}+\Gamma^{\mu}{}_{\nu\lambda}T^{\lambda\nu}+\Gamma^{\nu}{}_{\nu\lambda}T^{\mu\lambda}=0.
\]
For $\mu=R$ we obtain
\[
p_{R,R}+(\psi+\nu+\mu_{\vartheta})_{,R}p_R+
\left[\nu_{,R}e^{2\nu}-(\omega-\Omega)\omega_{,R}e^{2\psi}-(\omega-\Omega)^2\psi_{,R}e^{2\psi}\right]
\frac{\rho}{V^2}
\]
\begin{equation}\label{hyd1}
+\left[-(\psi+\mu_{\vartheta})_{,R}e^{2\nu}-(\omega-\Omega)\omega_{,R}e^{2\psi}
+(\omega-\Omega)^2(\nu+\mu_\vartheta)_{,R}e^{2\psi}\right]\frac{p_\vartheta}{V^2}=0,
\end{equation}
whereas for $\mu=\vartheta$ we have
\[
p_{\vartheta,\vartheta}+\left[(\nu+\mu_R)_{,\vartheta}e^{2\nu}-(\omega-\Omega)\omega_{,\vartheta}e^{2\psi}
-(\omega-\Omega)^2(\psi+\mu_R)_{,\vartheta}e^{2\psi}\right]\frac{p_\vartheta}{V^2}
\]
\begin{equation}\label{hyd2}
+\left[\nu_{,\vartheta}e^{2\nu}-(\omega-\Omega)\omega_{,\vartheta}e^{2\psi}-(\omega-\Omega)^2\psi_{,\vartheta}e^{2\psi}\right]
\frac{\rho}{V^2}-\mu_{R,\vartheta}p_R=0.
\end{equation}
The set of eight equations (\ref{G1})-(\ref{hyd2}) for the eight
unknown functions $\nu$, $\psi$, $\mu_R$, $\mu_\vartheta$,
$\omega$, $\Omega$, $p_R$, and $p_\vartheta$ represents the full
set of Einstein field equations assuming an anisotropic
energy-momentum tensor. If the energy density is inspired by
Noncommutative Geometry in the spirit of
\cite{nico1,nico2,spallucci3,ansoldi1,nico3,spallucci2} and an
equation of state of the de Sitter form is assumed, namely
$\rho=-p_R$, it is not difficult to verify that equation
(\ref{hyd1}) becomes
\begin{equation}\label{generalization}
-p_\vartheta=\rho+\frac{V^2}{\Phi}~\rho_{,R},
\end{equation}
where
\[
\Phi=\Phi(R,\vartheta)=(\psi+\mu_\vartheta)_{,R}e^{2\nu}+(\omega-\Omega)\omega_{,R}e^{2\psi}
-(\omega-\Omega)^2(\nu+\mu_{\vartheta})_{,R}e^{2\psi}.
\]
Equation (\ref{generalization}) generalizes formula (41) in
\cite{spallucci2} to any stationary, axial-symmetric geometry
generated by a perfect anisotropic fluid with $\ell^\mu$ chosen as
in the present section.

\section{Einstein field equations with $\ell^t=\ell^\varphi=0$}
In the case of $\ell^\mu$ specified by relation (\ref{L2})
attention has to be paid to the contravariant component
$T^{R\vartheta}$ of the energy-momentum tensor, since it does not
vanish any longer and is now given by
\[
T^{R\vartheta}=\frac{1}{2}(p_R-p_\vartheta)e^{-(\mu_R+\mu_\vartheta)}.
\]
Moreover,
\[
T^{\vartheta\vartheta}=\frac{1}{2}(p_R+p_\vartheta)e^{-2\mu_\vartheta},\quad
T^{RR}=\frac{1}{2}(p_R+p_\vartheta)e^{-2\mu_R},
\]
whereas the remaining non-vanishing components are the same as
those given in the previous section. The corresponding
contravariant components in the non-coordinate basis read
\[
T^{(R)(R)}=T^{(\vartheta)(\vartheta)}=\frac{1}{2}(p_R+p_\vartheta),\quad
T^{(R)(\vartheta)}=\frac{1}{2}(p_R-p_\vartheta).
\]
Einstein field equations (\ref{G1})-(\ref{G3}) remain the same,
whereas  (\ref{G4})-(\ref{G6})become now
\[
(\psi+\nu)_{,R,\vartheta}-(\psi+\nu)_{,R}\mu_{R,\vartheta}-(\psi+\nu)_{,\vartheta}\mu_{\vartheta,R}
\]
\begin{equation}\label{G7}
+\psi_{,R}\psi_{,\vartheta}
+\nu_{,R}\nu_{,\vartheta}-\frac{1}{2}\omega_{,R}\omega_{,\vartheta}e^{2(\psi-\nu)}=4\pi(p_R-p_\vartheta)e^{-(\mu_R+\mu_\vartheta)},
\end{equation}
\[
e^{-2\mu_R}\left[\nu_{,R}\left(\psi+\mu_\vartheta\right)_{,R}+\psi_{,R}\mu_{\vartheta,R}\right]+
e^{-2\mu_\vartheta}\left[\left(\psi+\nu\right)_{,\vartheta,\vartheta}+(\psi+\nu)_{,\vartheta}(\nu-\mu_\vartheta)_{,\vartheta}
+\psi_{,\vartheta}\psi_{,\vartheta}\right]
\]
\begin{equation}\label{G8}
+\frac{1}{4}e^{2\psi-2\nu}\left[(\omega_{,R})^2
e^{-2\mu_R}-(\omega_{,\vartheta})^2
e^{-2\mu_\vartheta}\right]=-4\pi(p_{R}+p_\vartheta),
\end{equation}
\[
e^{-2\mu_R}\left[\left(\psi+\nu\right)_{,R,R}+(\psi+\nu)_{,R}(\nu-\mu_R)_{,R}+\psi_{,R}\psi_{,R}\right]+
e^{-2\mu_\vartheta}\left[\nu_{,\vartheta}\left(\psi+\mu_R\right)_{,\vartheta}+\psi_{,\vartheta}\mu_{R,\vartheta}\right]
\]
\begin{equation}\label{G9}
-\frac{1}{4}e^{2\psi-2\nu}\left[(\omega_{,R})^2
e^{-2\mu_R}-(\omega_{,\vartheta})^2
e^{-2\mu_\vartheta}\right]=-4\pi(p_R+p_\vartheta).
\end{equation}
In this case the hydrodynamic equations are
\[
(p_R+p_\vartheta)_{,R}+(\psi+\nu)_{,R}(p_R+p_\vartheta)+
2\left[\nu_{,R}e^{2\nu}-(\omega-\Omega)\omega_{,R}e^{2\psi}-(\omega-\Omega)^2\psi_{,R}e^{2\psi}\right]
\frac{\rho}{V^2}
\]
\[
+2\left[-(\psi+\mu_{\vartheta})_{,R}e^{2\nu}-(\omega-\Omega)\omega_{,R}e^{2\psi}
+(\omega-\Omega)^2(\nu+\mu_\vartheta)_{,R}e^{2\psi}\right]\frac{p_\vartheta}{V^2}
\]
\begin{equation}\label{hyd3}
+\left[(\nu+2\mu_R+\psi)_{,\vartheta}(p_R-p_\vartheta)+(p_R-p_\vartheta)_{,\vartheta}\right]e^{\mu_R-\mu_\vartheta}=0,
\end{equation}
whereas for $\mu=\vartheta$ we have
\[
(p_R+p_\vartheta)_{,\vartheta}+(\psi+\nu)_{,\vartheta}(p_R+p_\vartheta)+
2\left[\nu_{,\vartheta}e^{2\nu}-(\omega-\Omega)\omega_{,\vartheta}e^{2\psi}
-(\omega-\Omega)^2\psi_{,\vartheta}e^{2\psi}\right]
\frac{\rho}{V^2}
\]
\[
+2\left[-\psi_{,\vartheta}e^{2\nu}-(\omega-\Omega)\omega_{,\vartheta}e^{2\psi}
+(\omega-\Omega)^2\nu_{,\vartheta}e^{2\psi}\right]\frac{p_\vartheta}{V^2}
\]
\begin{equation}\label{hyd4}
+\left[(\nu+2\mu_\vartheta+\psi)_{,R}(p_R-p_\vartheta)+(p_R-p_\vartheta)_{,R}\right]e^{-\mu_R+\mu_\vartheta}=0.
\end{equation}

\section{Einstein field equations for Kerr-like metrics}
Let us suppose that the metric (\ref{sam}) admits an event
horizon, i.e. a smooth two-dimensional null surface spanned by the
Killing vectors $\partial_t$ and $\partial_\varphi$. Suppose
further that the equation of this horizon be $H(R,\vartheta)=0$.
Such a surface will be null if
\[
g^{\mu\nu}H_{,\mu}H_{,\nu}=0,
\]
that is
\begin{equation}\label{pimpa1}
e^{2(\mu_\vartheta-\mu_R)}(H_{,R})^2+(H_{,\vartheta})^2=0.
\end{equation}
Furthermore, the gauge freedom allows to suppose that
\begin{equation}\label{pimpa2}
e^{2(\mu_\vartheta-\mu_R)}=\Delta(R)
\end{equation}
and (\ref{pimpa1}) implies that the equation of the event horizon
is simply $\Delta(R)=0$. Moreover, the condition that the surface
describing the event horizon be spanned by the Killing vectors
$\partial_t$ and $\partial_\varphi$ requires that the determinant
of the metric of the subspace $(t,\varphi)$ vanish on
$\Delta(R)=0$, i.e. $e^{2(\psi+\nu)}=0$ on $\Delta(R)=0$. Without
loss of generality we can suppose that
\begin{equation}\label{pimpa3}
e^\beta=e^{\psi+\nu}=\sqrt{\Delta}F(R,\theta),
\end{equation}
where $F$ is some function of $R$ and $\vartheta$ that we demand
to be regular on $\Delta(R)=0$ and on the axis $\vartheta=0$. In
virtue of the previous considerations we can establish the
following lemma:
\begin{lemma}
Under the condition that there exists at least one horizon the
metric (\ref{sam}) can be rewritten as
\begin{equation}\label{pimpa4}
ds^2=e^{\beta}\left(\chi-\frac{\omega^2}{\chi}\right)dt^2+2\frac{\omega
e^\beta}{\chi}~dtd\varphi-\frac{e^{2\mu_\vartheta}}{\Delta}~dR^2-e^{2\mu_\vartheta}~d\vartheta^2-
\frac{e^\beta}{\chi}~d\varphi^2
\end{equation}
with $\chi=e^{-\psi+\nu}$.
\end{lemma}
Let us introduce the vierbein
\[
e^{(t)}=e^{\beta/2}\sqrt{\chi}~dt,\quad
e^{\varphi}=\frac{e^{\beta/2}}{\sqrt{\chi}}(d\varphi-\omega
dt),\quad e^{(R)}=\frac{e^{\mu_\vartheta}}{\sqrt{\Delta}},\quad
e^{\vartheta}=e^{\mu_\vartheta}.
\]
We consider the energy-momentum tensor given as in
(\ref{enmomten}). In the present case, the non-vanishing
contravariant components of the fluid velocity are
\[
u^t=e^{-\beta/2}\sqrt{\frac{\chi}{\chi^2-(\omega-\Omega)^2}},\quad
u^\varphi=\Omega u^t.
\]
with $\Omega$ depending in general on $R$ and $\vartheta$, whereas
the unit space-like vector $\ell^\mu$ can be chosen to be
\[
\ell^\mu=\sqrt{\Delta}e^{-\mu_\vartheta}\delta^\mu_R.
\]
Finally, the covariant components of the energy-momentum tensor in
the non-coordinate basis introduced above are
\[
T_{(t)(t)}=\frac{\chi^2\rho+(\omega-\Omega)^2
p_\vartheta}{\chi^2-(\omega-\Omega)^2},\quad
T_{(t)(\varphi)}=\frac{(\omega-\Omega)\chi
}{\chi^2-(\omega-\Omega)^2}~(\rho+p_\vartheta),
\]
\[
T_{(\varphi)(\varphi)}=\frac{(\omega-\Omega)^2\rho+\chi^2
p_\vartheta}{\chi^2-(\omega-\Omega)^2},\quad T_{(R)(R)}=p_R,\quad
T^{(\vartheta)(\vartheta)}=p_\vartheta.
\]
Following \cite{chandra}, it results convenient to consider
Einstein field equations as taken below
\begin{equation}\label{A}
R_{(t)(t)}-R_{(\varphi)(\varphi)}=8\pi(T-T_{(t)(t)}+T_{(\varphi)(\varphi)}),
\end{equation}
\begin{equation}\label{B}
R_{(t)(t)}+R_{(\varphi)(\varphi)}=-8\pi(T_{(t)(t)}+T_{(\varphi)(\varphi)}),
\end{equation}
\begin{equation}\label{C}
R_{(t)(\varphi)}=-8\pi T_{(t)(\varphi)},
\end{equation}
\begin{equation}\label{D}
R_{(R)(\vartheta)}=0,
\end{equation}
\begin{equation}\label{E}
G_{(R)(R)}-G_{(\vartheta)(\vartheta)}=-8\pi(T_{(R)(R)}-T_{(\vartheta)(\vartheta)}).
\end{equation}
\begin{equation}\label{F}
G_{(R)(R)}+G_{(\vartheta)(\vartheta)}=-8\pi(T_{(R)(R)}+T_{(\vartheta)(\vartheta)}),
\end{equation}
The reason for this is twofold. First, it permits to express
Einstein field equations in a compact form in terms of the unknown
functions appearing in the metric (\ref{pimpa4}). Second, when we
rewrite the above equations in terms of the unknown functions
$\beta$, $\Delta$, $\chi$, $\omega$ and $\mu_\vartheta$, it
results that equation (\ref{F}) coincides with equation (\ref{A})
showing that not all Einstein field equations (\ref{A})-(\ref{F})
are independent. For our present purposes it is sufficient to
consider equations (\ref{A})-(\ref{E}), which now read
\begin{equation}\label{EFE1}
\left(\sqrt{\Delta}(\sqrt{\Delta}F)_{,R}\right)_{,R}+F_{,\vartheta,\vartheta}=-8\pi
Fe^{2\mu_\vartheta}(p_R+p_\vartheta),
\end{equation}
\begin{equation}\label{EFE2}
(\Delta
F(\ln{\chi})_{,R})_{,R}+(F(\ln{\chi})_{,\vartheta})_{,\vartheta}
-\frac{F}{\chi^2}\left[\Delta(\omega_{,R})^2+(\omega_{,\vartheta})^2\right]=
-8\pi fFe^{2\mu_\vartheta}(\rho+p_\vartheta),
\end{equation}
\begin{equation}\label{EFE3}
\left(\frac{\Delta
F}{\chi^2}~\omega_{,R}\right)_{,R}+\left(\frac{F}{\chi^2}~\omega_{,\vartheta}\right)_{,\vartheta}=
-16\pi gFe^{2\mu_\vartheta}(\rho+p_\vartheta),
\end{equation}
\[
(\ln{F})_{,R,\vartheta}-\left[(\ln{(\sqrt{\Delta}F)})_{,R}\mu_{\vartheta,\vartheta}+(\ln{F})_{,\vartheta}\mu_{\vartheta,R}\right]+
\frac{1}{2}(\ln{(\sqrt{\Delta}F)})_{,R}(\ln{F})_{,\vartheta}
\]
\begin{equation}\label{EFE4}
+\frac{\chi_{,R}\chi_{,\vartheta}-\omega_{,R}\omega_{,\vartheta}}{2\chi^2}=0,
\end{equation}
\[
\left(\sqrt{\Delta}(\sqrt{\Delta}F)_{,R}\right)_{,R}-F_{,\vartheta,\vartheta}+2F\left[
(\ln{F})_{,\vartheta}\left(\frac{1}{4}(\ln{F})_{,\vartheta}+\mu_{\vartheta,\vartheta}\right)
-\frac{1}{4}\left((\ln{\chi})_{,\vartheta}\right)^2 \right]
\]
\[
-2\Delta F\left[
(\ln{(\sqrt{\Delta}F)})_{,R}\left(\frac{1}{4}(\ln{(\sqrt{\Delta}F)})_{,R}+\mu_{\vartheta,R}\right)
-\frac{1}{4}\left((\ln{\chi})_{,R}\right)^2 \right]
\]
\begin{equation}\label{EFE5}
-\frac{F}{2\chi^2}\left[\Delta(\omega_{,R})^2-(\omega_{,\vartheta})^2\right]=8\pi
Fe^{2\mu_{\vartheta}}(p_R-p_\vartheta)
\end{equation}
with
\[
f(\chi,\omega,\Omega)=\frac{\chi^2+(\omega-\Omega)^2}{\chi^2-(\omega-\Omega)^2},\quad
g(\chi,\omega,\Omega)=\frac{\omega-\Omega}{\chi^2-(\omega-\Omega)^2},
\]
whereas the conservation equation for the energy-momentum tensor
gives rise to the following equations
\begin{equation}\label{conservazione1}
p_{R,R}+\left(\beta+\mu_\vartheta\right)_{,R}p_R
+\frac{f_{+}\rho+f_{-}p_\vartheta}{2\chi[\chi^2-(\omega-\Omega)^2]}=0,
\end{equation}
\begin{equation}\label{conservazione2}
p_{\vartheta,\vartheta}-\mu_{\vartheta,\vartheta}p_R
+\frac{g_{+}\rho+g_{-}p_\vartheta}{2\chi[\chi^2-(\omega-\Omega)^2]}=0,
\end{equation}
where $\beta=\ln{(\sqrt{\Delta}F)}$ and
\[
f_+=f_+(R,\vartheta)=\chi^2(\chi\beta_{,R}+\chi_{,R})-2(\omega-\Omega)\chi\omega_{,R}-(\omega-\Omega)^2(\chi\beta_{,R}-\chi_{,R})
,
\]
\[
f_-=f_{-}(R,\vartheta)=-\chi^2(\chi\beta_{,R}-\chi_{,R}+2\mu_{\vartheta,R}\chi)-2(\omega-\Omega)\chi\omega_{,R}
\]
\[
+(\omega-\Omega)^2(\chi\beta_{,R}+\chi_{,R}+2\mu_{\vartheta,R}\chi),
\]
\[
g_+=g_+(R,\vartheta)=\chi^2(\chi\beta_{,\vartheta}+\chi_{,\vartheta})-2(\omega-\Omega)\chi\omega_{,\vartheta}
-(\omega-\Omega)^2(\chi\beta_{,\vartheta}-\chi_{,\vartheta}),
\]
\[
g_-=g_{-}(R,\vartheta)=\chi^2(\chi\beta_{,\vartheta}+\chi_{,\vartheta}+2\mu_{\vartheta,\vartheta}\chi)
-2(\omega-\Omega)\chi\omega_{,\vartheta}
\]
\[
-(\omega-\Omega)^2(\chi\beta_{,\vartheta}-\chi_{,\vartheta}+2\mu_{\vartheta,\vartheta}\chi).
\]
A this point a couple of remarks are in order. First of all, as
expected equation (\ref{EFE1}) reduces to equation (43) at page
$279$ in \cite{chandra} whenever $F$ depends only on the angular
variable $\vartheta$. In this case one recovers the well known
result for the Kerr metric given by $\Delta(R)=R^2-2MR+a^2$.
Furthermore, the system of five equations
(\ref{EFE1})-(\ref{EFE5}) contains 8 unknowns, namely five metric
functions and the mass density, the radial pressure and the
tangential pressure, which until now we left unspecified. To
circumvent this problem we recall that we are interested in
Noncommutative Geometry inspired Kerr-like solutions and hence,
the mass density $\rho$ is an assigned function and an equation of
state of the de Sitter form, namely $p_R=-\rho$ will be assumed
following the line of reasoning in
\cite{nico1,nico2,spallucci3,ansoldi1,nico3,spallucci2}. Finally,
the pressures $p_\vartheta$ and $p_\varphi$ can be computed by
invoking the conservation equation for the energy-momentum tensor.
In this setting (\ref{conservazione1}) reduces as expected to the
equation
\begin{equation}\label{pressa}
-p_\vartheta=\rho+U\rho_{,R},
\end{equation}
where
\[
U=U(R,\vartheta)=-\frac{2\chi}{f_{-}}~[\chi^2-(\omega-\Omega)^2].
\]
Finally, the next theorem shows that the metric (44) given in
\cite{spallucci2} can never satisfy Einstein Field equations
(\ref{EFE1})-(\ref{EFE5}).
\begin{theorem}\label{teorema elegante}
Under the condition that there exists at least one horizon the
Einstein field equations (\ref{EFE1})-(\ref{EFE5}) with
energy-momentum tensor given by
\[
T^{\mu}{}_{\nu}=(\rho+p_\vartheta)(u^\mu
u_\nu-\ell^\mu\ell_\nu)-p_\vartheta\delta^\mu_\nu,\quad\ell^\mu=e^{-\mu_\vartheta}\sqrt{\Delta}~\delta^\mu_R,\quad
\ell^\vartheta=0
\]
with
\begin{equation}\label{densita-rho}
\rho(R,\theta)=\frac{R^4}{\Sigma^2}\rho_G(R),\quad\Sigma=R^2+a^2\cos^2{\vartheta},\quad
\rho_G(R)=\frac{M}{(4\pi\ell_0)^{3/2}}e^{-R^2/4\ell_0^2}
\end{equation}
do not admit any solution of the form
\[
ds^2=\left(1-\frac{2RM(R)}{\Sigma}\right)dt^2+\frac{4aRM(R)}{\Sigma}\sin^2{\vartheta}dtd\varphi-\frac{\Sigma}{\Delta}dR^2
-\Sigma d\vartheta^2
\]
\begin{equation}\label{spal-met}
-\frac{\sin^2{\vartheta}}{\Sigma}\left[(R^2+a^2)^2-a^2\sin^2{\vartheta}\Delta\right]d\varphi^2,
\end{equation}
where
\begin{equation}\label{deltozzo}
\Delta=R^2-2RM(R)+a^2,\quad M(R)=4\pi\int_0^R dx~x^2\rho_G(x).
\end{equation}
\end{theorem}
\begin{Proof.}
It sufficient to show that the Einstein field equation
(\ref{EFE1}) will never be satisfied by the metric
(\ref{spal-met}). First of all, comparing the metrics
(\ref{pimpa4}) and (\ref{spal-met}) we find that our function $F$
has to depend only on the angular variable $\vartheta$ and more
precisely, $F(\vartheta)=\sin{\vartheta}$. Notice that the same
choice occurs in the derivation of the classic Kerr metric
\cite{chandra}. Moreover, the function $\Delta$ can be identified
with $\Delta=R^2-2RM(R)+a^2$ and $e^{2\mu_\vartheta}=\Sigma$. If
we consider an equation of state of the form $p_R=-\rho$ as in
\cite{spallucci2}, equation (\ref{EFE1}) simplifies as follows
\begin{equation}\label{EFE11}
\frac{d^2\Delta}{dR^2}-2=16\pi\Sigma(\rho-p_\vartheta).
\end{equation}
Taking into account equation (41) in \cite{spallucci2} or
equivalently our (\ref{pressa}), the r.h.s. of (\ref{EFE11}) can
be written in terms of $\rho$ and its derivative with respect to
the space-like variable $R$. Thus, we find the equation
\begin{equation} \label{m1}
\frac{d^2\Delta}{dR^2}=2+8\left(4\pi R^2\rho_G+\pi
R^3\frac{d\rho_G}{dR}\right),
\end{equation}
whose solution is
\[
\Delta(R)=R^2+C_1
R+C_2+\frac{M}{2}\sqrt{2\ell_0}R\left[\ell_0~{\rm{erf}}\left(\frac{R}{2\ell_0}\right)-\frac{R}{\sqrt{\pi}}~e^{-R^2/4\ell_0}\right].
\]
But $\Delta(R)$ was already given by (\ref{deltozzo}) whose
integration does not agree with the above expression. Therefore,
we have a contradiction. ~~$\square$
\end{Proof.}
We can sharpen the above statement by noticing that alone from
(\ref{deltozzo}) we obtain
\[
\frac{d^2\Delta}{dR^2}=2-8\left(4\pi R^2\rho_G+\pi
R^3\frac{d\rho_G}{dR}\right),
\]
which is not the same  as (\ref{m1}). Hence, we can assert
\begin{corollary}
The statement of Theorem 1 is independent of the choice of the
density $\rho$.
\end{corollary}
Notice that the Theorem 1 has been proved by assuming the
existence of a horizon. However, this is not necessary. Indeed,
with the correct choice of the four-velocities, i.e,
$u^\mu=\frac{R^2+a^2}{\sqrt{\Sigma\Delta}}\left(\delta_t^\mu+\Omega\delta_\varphi^\mu\right),
\Omega=\frac{a}{R^2+a^2}$ we can prove the following:
\begin{theorem}\label{teoremozzo}
The Einstein field equations
\[
G_{\mu\nu}=-8\pi T_{\mu\nu}
\]
with energy-momentum tensor given by
\begin{equation}\label{tensore en mom1}
T^{\mu}{}_{\nu}=(\rho+p_\vartheta)(u^\mu
u_\nu-\ell^\mu\ell_\nu)-p_\vartheta\delta^\mu_\nu
\end{equation}
and
\begin{equation}\label{componentinuove}
u^\mu=\frac{R^2+a^2}{\sqrt{\Sigma\Delta}}\left(\delta_t^\mu+\Omega\delta_\varphi^\mu\right),\quad
\frac{u^\varphi}{u^t}=\Omega=\frac{a}{R^2+a^2},\quad
\ell^\mu=-\sqrt{\frac{\Delta}{\Sigma}}~\delta_{R}^\mu
\end{equation}
do not admit any solution of the form
\[
ds^2=\left(1-\frac{2RM(R)}{\Sigma}\right)dt^2+\frac{4aRM(R)}{\Sigma}\sin^2{\vartheta}dtd\varphi-\frac{\Sigma}{\Delta}dR^2
-\Sigma d\vartheta^2
\]
\begin{equation}\label{metrica11}
-\frac{\sin^2{\vartheta}}{\Sigma}\left[(R^2+a^2)^2-a^2\sin^2{\vartheta}\Delta\right]d\varphi^2
\end{equation}
with
\[
\Sigma=R^2+a^2\cos^2{\vartheta},\quad\Delta=R^2-2RM(R)+a^2.
\]
\end{theorem}
\begin{Proof.}
We begin by studying the conservation equation
$T^{\mu\nu}{}_{;\nu}=0$. For $\mu=R$ we obtain
\begin{equation}\label{componente
radiale}
\partial_R
T^{RR}+\Gamma^{R}_{tt}T^{tt}+2\Gamma^{R}_{t\varphi}T^{t\varphi}+\Gamma^{R}_{\varphi\varphi}T^{\varphi\varphi}
+(\Gamma^{t}_{tR}+\Gamma^{\varphi}_{\varphi
R}+2\Gamma^{R}_{RR}+\Gamma^{\vartheta}_{\vartheta
R})T^{RR}+\Gamma^{R}_{\vartheta\vartheta}T^{\vartheta\vartheta}=0.
\end{equation}
Taking into account that
\[
T^{RR}=-\frac{\Delta}{\Sigma}~\rho,\quad
T^{tt}=\frac{(R^2+a^2)^2}{\Sigma\Delta}~\rho+\frac{a^2\sin^2{\vartheta}}{\Sigma}~p_\vartheta,\quad
T^{t\varphi}=\frac{a(R^2+a^2)}{\Sigma\Delta}~\rho+\frac{a}{\Sigma}~p_\vartheta
\]
\[
T^{\varphi\varphi}=\frac{a^2}{\Sigma\Delta}~\rho+\frac{1}{\Sigma\sin^{2}{\vartheta}}~p_\vartheta,\quad
T^{\vartheta\vartheta}=\frac{p_\vartheta}{\Sigma}
\]
and
\[
\Gamma^{R}_{tt}=-\frac{\Delta}{\Sigma^3}\left[R\Sigma\frac{dM}{dR}-(R^2-a^2\cos^2{\vartheta})M(R)\right],\quad
\Gamma^{R}_{t\varphi}=-a\sin^2{\vartheta}~\Gamma^{R}_{tt},
\]
\[
\Gamma^{R}_{\varphi\varphi}=\frac{\sin^2{\vartheta}}{\Sigma}(a^2\Sigma\sin^2{\vartheta}~\Gamma^{R}_{tt}-R\Delta),\quad
\Gamma^{t}_{tR}=\frac{(R^2+a^2)\Sigma}{\Delta^2}~\Gamma^{R}_{tt},
\]
\[
\Gamma^{\varphi}_{\varphi
R}=-\frac{1}{\Sigma^2\Delta}\left[-Ra^2\Sigma\sin^{2}{\vartheta}\frac{dM}{dR}+a^2(R^2-a^2\cos^2{\vartheta})M(R)\right.
\]
\[
\left.+a^2\Sigma\cos^2{\vartheta}M(R)-a^2
R\cos^2{\vartheta}(2R^2+a^2\cos^2{\vartheta})-R^4(R-2M(R))\right],
\]
\[
\Gamma^{R}_{RR}=\frac{1}{\Sigma\Delta}\left(a^2
R\sin^2{\vartheta}-\frac{\Sigma^3}{\Delta}~\Gamma^{R}_{tt}\right),\quad
\Gamma^{\vartheta}_{\vartheta R}=\frac{R}{\Sigma},\quad
\Gamma^{R}_{\vartheta\vartheta}=-\frac{R\Delta}{\Sigma},
\]
\[
\Gamma^{t}_{tR}+\Gamma^{\varphi}_{\varphi
R}+2\Gamma^{R}_{RR}+\Gamma^{\vartheta}_{\vartheta R}=\frac{1}{
\Sigma\Delta}\left[-\frac{\Sigma^3}{\Delta}~\Gamma^{R}_{tt}+2R^2(R-2M(R))+a^2
R(3-\cos^2{\vartheta})\right],
\]
after a tedious manipulation equation (\ref{componente radiale})
becomes
\[
A(R,\vartheta)~p_\vartheta=B(R,\vartheta)~\rho+\frac{\Delta}{\Sigma}~\partial_R\rho
\]
with
\[
A(R,\vartheta)=\frac{1}{\Sigma}\left(a^2\sin^2{\vartheta}~\Gamma^{R}_{tt}+2a~\Gamma^{R}_{t\varphi}
+\frac{\Gamma^{R}_{\varphi\varphi}}{\sin^2{\vartheta}}+\Gamma^{R}_{\vartheta\vartheta}\right)=-\frac{2R\Delta}{\Sigma^2}
\]
and
\[
B(R,\vartheta)=\partial_R\left(\frac{\Delta}{\Sigma}\right)-\frac{(R^2+a^2)^2}{\Sigma\Delta}~\Gamma^{R}_{tt}
-\frac{2a(R^2+a^2)}{\Sigma\Delta}~\Gamma^{R}_{t\varphi}-\frac{a^2}{\Sigma\Delta}~\Gamma^{R}_{\varphi\varphi}
\]
\[
+\frac{\Delta}{\Sigma}(\Gamma^{t}_{tR}+\Gamma^{\varphi}_{\varphi
R}+2\Gamma^{R}_{RR}+\Gamma^{\vartheta}_{\vartheta
R})=\frac{2R\Delta}{\Sigma^2}.
\]
Finally, equation (\ref{componente radiale}) reduces to
\begin{equation}\label{primaeq}
-p_\theta=\rho+\frac{\Sigma}{2R}~\partial_R\rho
\end{equation}
which coincides with equation (41) in \cite{spallucci2}. For the
case $\mu=\vartheta$ the conservation equation for the
energy-momentum tensor gives rise to the following equation
\begin{equation}\label{componente
angolare}
\partial_\vartheta
T^{\vartheta\vartheta}+\Gamma^{\vartheta}_{tt}T^{tt}+2\Gamma^{\vartheta}_{t\varphi}T^{t\varphi}+
\Gamma^{\vartheta}_{\varphi\varphi}T^{\varphi\varphi}
+(\Gamma^{t}_{t\vartheta}+\Gamma^{R}_{R\vartheta
}+2\Gamma^{\vartheta}_{\vartheta\vartheta}+\Gamma^{\varphi}_{\vartheta\varphi
})T^{\vartheta\vartheta}+\Gamma^{\vartheta}_{RR}T^{RR}=0,
\end{equation}
where the Christoffel symbols are given by
\[
\Gamma^{\vartheta}_{t\varphi}=\frac{2aR}{\Sigma^3}(R^2+a^2)M(R)\sin{\vartheta}\cos{\vartheta},\quad
\Gamma^{\vartheta}_{tt}=-\frac{a}{R^2+a^2}~\Gamma^{\vartheta}_{t\varphi},
\]
\[
\Gamma^{\vartheta}_{\varphi\varphi}=-\frac{\sin{\vartheta}\cos{\vartheta}}{\Sigma^3}
\left[\Sigma^2\Delta+2R(R^2+a^2)^2M(R)\right] ,\quad
\Gamma^{\vartheta}_{RR}=\frac{a^2\sin{\vartheta}\cos{\vartheta}}{\Sigma\Delta},
\]
\[
\Gamma^{t}_{t\vartheta}=-\frac{a\Sigma}{R^2+a^2}~\Gamma^{\vartheta}_{t\varphi},\quad
\Gamma^{R}_{R\vartheta}=\Gamma^{\vartheta}_{\vartheta\vartheta}=-\Gamma^{\vartheta}_{RR},\quad
\Gamma^{\varphi}_{\varphi\vartheta}=\frac{\cot{\vartheta}}{\Sigma}\left(\Sigma^2+2a^2
RM(R)\sin^2{\vartheta}\right)
\]
\[
\Gamma^{t}_{t\vartheta}+\Gamma^{R}_{R\vartheta
}+2\Gamma^{\vartheta}_{\vartheta\vartheta}+\Gamma^{\varphi}_{\vartheta\varphi
}=\frac{\cot{\vartheta}}{\Sigma}\left(R^2+4a^2\cos^2{\vartheta}
-3a^2\right).
\]
After a long computation equation (\ref{componente angolare}) can
be written in the more amenable form
\begin{equation}\label{angolare2}
\partial_\vartheta
p_\vartheta-\frac{2a^2}{\Sigma}\sin{\vartheta}\cos{\vartheta}(\rho+p_\vartheta)=0.
\end{equation}
Notice that in the limit $a\to 0$ the above equation reduces to
$\partial_\vartheta p_\vartheta$ as it is expected for the
noncommutative inspired Schwarzschild metric. Employing
(\ref{primaeq}) the above equation becomes a partial differential
equation for the function $\rho(R,\vartheta)$, namely
\begin{equation}\label{PDE}
\partial_\vartheta\rho-\frac{a^2}{R}\sin{\vartheta}\cos{\vartheta}~\partial_R\rho
+\frac{1}{2R}~\partial_\vartheta(\Sigma~\partial_R\rho)=0.
\end{equation}
The above equation can be solved by the method of separation of
variables. To this purpose, let
$\rho(R,\vartheta)=F_1(R)F_2(\vartheta)$. Then, we obtain
\begin{equation}\label{F1}
\dot{F}_1=\frac{1}{RF_1}\left(2R{\dot{F}_1}^2+F_1\dot{F}_1\right)
\end{equation}
and
\begin{equation}\label{F2}
\frac{dF_2}{d\vartheta}=\frac{2a^2\sin{\vartheta}\cos{\vartheta}~\dot{F}_1}{2RF_1+\Sigma\dot{F}_1
}~F_2,
\end{equation}
where the dot denotes differentiation with respect to the variable
$R$. The solution of the first order non linear equation
(\ref{F1}) is given by
\begin{equation}\label{solF1}
F_1(R)=-\frac{2}{C_1 R^2+C_2}.
\end{equation}
Substituting (\ref{solF1}) into (\ref{F2}) we get
\[
\frac{dF_2}{d\vartheta}=\frac{2C_1
a^2\sin{\vartheta}\cos{\vartheta}}{a^2
C_1\cos^2{\vartheta}-2C_2}~F_2,
\]
whose solution is
\begin{equation}\label{solutio_2}
F_2(\vartheta)=-\frac{2C_3}{a^2 C_1\cos{2\vartheta}+a^2 C_1-4C_2}.
\end{equation}
On the other side, taking into account that
$T_{RR}=-(\Sigma/\Delta)~\rho$ and making use of (\ref{comp_RR})
the Einstein field equation $G_{RR}=-8\pi T_{RR}$ simplifies to
\[
\frac{dM}{dR}=4\pi\frac{\Sigma^2}{R^2}\rho
\]
with $\rho(R,\vartheta)=F_1(R)F_2(\vartheta)$ where the functions
$F_1(R)$ and $F_2(\vartheta)$ are given by (\ref{solF1}) and
(\ref{solutio_2}). Since the mass function depends only on the
spatial variable $R$, whereas the r.h.s. of the above equation
manifests an additional dependence on the angular variable
$\vartheta$ which cannot be removed. We have a contradiction and
the proof is completed.~~$\square$
\end{Proof.}

\section{Conclusions}
The present work is in the spirit of
\cite{nico1,nico2,spallucci3,ansoldi1,nico3,spallucci2} and deals
with the problem of deriving the full set of Einstein field
equations assuming an anisotropic energy-momentum tensor, where
the energy density is inspired by Noncommutative Geometry and is
meant to replace the ring singularity. These equations are written
down in an explicit form in (\ref{G1})-(\ref{hyd2}) and
(\ref{EFE1})-(\ref{conservazione2}) if we demand the existence of
at least one horizon.. Any candidate solution of the Kerr metric
inspired by Noncommutative Geometry has to satisfy these
equations. A suggestion for such a solution was made in
(\cite{spallucci2}) where the metric is written in the Kerr-Schild
form in terms of a null vector $k^{\mu}$. This vector is chosen
explicitly on the basis of symmetry arguments such that only one
function remains unknown. For this function a differential
equation is derived out of the Einstein equations. It is, however,
a priori not clear if \textit{all} Einstein equations are
identically satisfied. We did this check and showed in theorem 1,
theorem 2 and appendix B that the so-called Kerrr solution can
never satisfy the Einstein field equations.  In this sense, our
results can be seen as complementary to (\cite{spallucci2}). We
hope that in the future using our set of equations we will be able
to examine closer the smeared ring singularity of the Kerr
solution inspired by Noncommutative Geometry.

\section{Appendix}
\section{List of the relevant Christoffel symbols and tensor components}
\subsection{Christoffel symbols}
For the metric (\ref{sam}) the non vanishing Christoffel symbols
are
\[
\Gamma^{t}{}_{ti}=\nu_{,i}-\frac{1}{2}\omega\omega_{,i}e^{2(\psi-\nu)},\quad
\Gamma^{t}{}_{\varphi i}=\frac{1}{2}\omega_{,i}e^{2(\psi-\nu)},
\]
\[
\Gamma^{\varphi}{}_{ti}=\omega\nu_{,i}-\frac{1}{2}\omega_{,i}-\omega\psi_{,i}
-\frac{1}{2}\omega^2\omega_{,i}e^{2(\psi-\nu)},\quad
\Gamma^{\varphi}{}_{\varphi
i}=\psi_{,i}+\frac{1}{2}\omega\omega_{,i}e^{2(\psi-\nu)},
\]
\[
\Gamma^{R}{}_{tt}=\nu_{,R}e^{2(\nu-\mu_R)}-(\omega\omega_{,R}+\omega^2\psi_{,R})e^{2(\psi-\mu_R)},\quad
\Gamma^{R}{}_{t\varphi}=\left(\frac{1}{2}\omega_{,R}+\omega\psi_{,R}\right)e^{2(\psi-\mu_R)},
\]
\[
\Gamma^{R}{}_{\varphi\varphi}=-\psi_{,R}e^{2(\psi-\mu_R)},\quad
\Gamma^{R}{}_{RR}=\mu_{R,R},\quad
\Gamma^{R}{}_{R\vartheta}=\mu_{R,\vartheta},\quad
\Gamma^{R}{}_{\vartheta\vartheta}=-\mu_{\vartheta,R}e^{2(\mu_\vartheta-\mu_R)},
\]
\[
\Gamma^{\vartheta}{}_{tt}=\nu_{,\vartheta}e^{2(\nu-\mu_\vartheta)}-(\omega\omega_{,\vartheta}
+\omega^2\psi_{,\vartheta})e^{2(\psi-\mu_\vartheta)},\quad
\Gamma^{\vartheta}{}_{t\varphi}=\left(\frac{1}{2}\omega_{,\vartheta}+\omega\psi_{,\vartheta}\right)e^{2(\psi-\mu_\vartheta)},
\]
\[
\Gamma^{\vartheta}{}_{\varphi\varphi}=-\psi_{,\vartheta}e^{2(\psi-\mu_\vartheta)},\quad
\Gamma^{\vartheta}{}_{RR}=-\mu_{R,\vartheta}e^{2(\mu_R-\mu_\vartheta)},\quad
\Gamma^{\vartheta}{}_{R\vartheta}=\mu_{\vartheta,R},\quad
\Gamma^{\vartheta}{}_{\vartheta\vartheta}=\mu_{\vartheta,\vartheta},
\]
where $i=R,\vartheta$.
\subsection{Covariant Riemann tensor components}
The non-vanishing components of the covariant Riemann tensor have
been computed according to the formula
\[
R_{\rho\lambda\mu\nu}=\frac{1}{2}
\left(g_{\lambda\mu,\rho\nu}-g_{\rho\mu,\lambda\nu}-g_{\lambda\nu,\rho\mu}+g_{\rho\nu,\lambda\mu}\right)+
g_{\xi\eta}\left(\Gamma^{\xi}{}_{\rho\nu}\Gamma^{\eta}{}_{\lambda\mu}-
\Gamma^{\xi}{}_{\rho\mu}\Gamma^{\eta}{}_{\lambda\nu}\right).
\]
For the metric (\ref{sam}) the relevant components of the
covariant curvature tensor are
\[
R_{\varphi t\varphi
t}=e^{2(\psi+\nu)}\left[-e^{-2\mu_R}\nu_{,R}\psi_{,R}-e^{-2\mu_\vartheta}\nu_{,\vartheta}\psi_{,\vartheta}
-\frac{1}{4}e^{2(\psi-\nu)}\left[
e^{-2\mu_R}(\omega_{,R})^2+e^{-2\mu_\vartheta}(\omega_{,\vartheta})^2\right]\right],
\]
\[
R_{RtRt}=e^{2(\nu+\mu_R)}\left[-e^{-\mu_R-\nu}\left(e^{\nu-\mu_R}\nu_{,R}\right)_{,R}-\nu_{,\vartheta}\mu_{R,\vartheta}e^{-2\mu_\vartheta}
+\frac{3}{4}(\omega_{,R})^2e^{2\psi-2\mu_R-2\nu}\right]
\]
\[
+\left[\omega\omega_{,R,R}+3\omega\omega_{,R}\psi_{,R}+\omega^2\psi_{,R,R}+\left(\omega\psi_{,R}\right)^2-\omega\omega_{,R}\nu_{,R}-\omega\omega_{,R}\mu_{R,R}-\omega^2\psi_{,R}\mu_{R,R}\right]e^{2\psi}
\]
\[
+\left[\mu_{R,\vartheta}\omega\omega_{,\vartheta}+\mu_{R,\vartheta}\omega^2\psi_{,\vartheta}\right]e^{2\psi-2\mu_\vartheta+2\mu_R}+\frac{1}{4}(\omega\omega_{,R})^2
e^{4\psi-2\nu},
\]
\[
R_{R t
R\varphi}=\left[-\frac{1}{2}\omega_{,R,R}-\frac{3}{2}\omega_{,R}\psi_{,R}-\omega\psi_{,R,R}-\omega\psi_{,R}^2
+\frac{1}{2}\omega_{,R}\nu_{,R}+\frac{1}{2}\omega_{,R}\mu_{R,R}+\omega\psi_{,R}\mu_{R,R}\right]e^{2\psi}
\]
\[
-\frac{1}{4}\omega\omega_{,R}^2
e^{4\psi-2\nu}-\mu_{R,\vartheta}e^{2\psi-2\mu_\vartheta+2\mu_R}\left[\frac{1}{2}\omega_{,\vartheta}+\omega\psi_{,\vartheta}\right]
=R_{R \varphi R t},
\]
\[
R_{R \varphi R
\varphi}=\left[\psi_{,R,R}+\psi_{,R}^2-\mu_{R,R}\psi_{,R}\right]e^{2\psi}
+\frac{1}{4}\omega_{,R}^2e^{4\psi-2\nu}+\mu_{R,\vartheta}\psi_{,\vartheta}e^{2\psi+2\mu_R-2\mu_\vartheta},
\]
\[
R_{\vartheta t \vartheta
t}=e^{2(\nu+\mu_\vartheta)}\left[-e^{-\mu_\vartheta-\nu}\left(e^{\nu-\mu_\vartheta}\nu_{,\vartheta}\right)_{,\vartheta}-\nu_{,R}\mu_{\vartheta,R}e^{-2\mu_R}
+\frac{3}{4}(\omega_{,\vartheta})^2e^{2\psi-2\mu_\vartheta-2\nu}\right]
\]
\[
+\left[\omega\omega_{,\vartheta,\vartheta}+3\omega\omega_{,\vartheta}\psi_{,\vartheta}+\omega^2\psi_{,\vartheta,\vartheta}+\left(\omega\psi_{,\vartheta}\right)^2-\omega\omega_{,\vartheta}\nu_{,\vartheta}-\omega\omega_{,\vartheta}\mu_{\vartheta,\vartheta}-\omega^2\psi_{,\vartheta}\mu_{\vartheta,\vartheta}\right]e^{2\psi}
\]
\[
+\left[\mu_{\vartheta,R}\omega\omega_{,R}+\mu_{\vartheta,R}\omega^2\psi_{,R}\right]e^{2\psi-2\mu_R+2\mu_\vartheta}+\frac{1}{4}(\omega\omega_{,\vartheta})^2
e^{4\psi-2\nu},
\]
\[
R_{\vartheta t \vartheta
\varphi}=\left[-\frac{1}{2}\omega_{,\vartheta,\vartheta}-\frac{3}{2}\omega_{,\vartheta}\psi_{,\vartheta}-\omega\psi_{,\vartheta,\vartheta}-\omega\psi_{,\vartheta}^2
+\frac{1}{2}\omega_{,\vartheta}\nu_{,\vartheta}+\frac{1}{2}\omega_{,\vartheta}\mu_{\vartheta,\vartheta}+\omega\psi_{,\vartheta}\mu_{\vartheta,\vartheta}\right]e^{2\psi}
\]
\[
-\frac{1}{4}\omega\omega_{,\vartheta}^2
e^{4\psi-2\nu}-\mu_{\vartheta,R}e^{2\psi-2\mu_R+2\mu_\vartheta}\left[\frac{1}{2}\omega_{,R}+\omega\psi_{,R}\right]
=R_{\vartheta t \vartheta\varphi},
\]
\[
R_{\vartheta \varphi \vartheta
\varphi}=\left[\psi_{,\vartheta,\vartheta}+\psi_{,\vartheta}^2-\mu_{\vartheta,\vartheta}\psi_{,\vartheta}\right]e^{2\psi}
+\frac{1}{4}\omega_{,\vartheta}^2e^{4\psi-2\nu}+\mu_{\vartheta,R}\psi_{,R}e^{2\psi+2\mu_\vartheta-2\mu_R},
\]
\[
R_{t R t
\vartheta}=e^{2\nu+\mu_R+\mu_\vartheta}\left[-e^{-\mu_\vartheta-\nu}\left(e^{\nu-\mu_R}\nu_{,R}\right)_{,\vartheta}+\nu_{,\vartheta}\mu_{\vartheta,R}e^{-\mu_R-\mu_\vartheta}
+\frac{3}{4}\omega_{,R}\omega_{,\vartheta}e^{2\psi-2\nu-\mu_R-\mu_\vartheta}\right]
\]
\[
+\left[\omega\omega_{,R,\vartheta}+\frac{3}{2}\omega\omega_{,R}\psi_{,\vartheta}+\frac{3}{2}\omega\omega_{,\vartheta}\psi_{,R}+
\omega^2\psi_{,R,\vartheta}+
\omega^2\psi_{,R}\psi_{,\vartheta}-\frac{1}{2}\omega\omega_{,R}\nu_{,\vartheta}-\frac{1}{2}\omega\omega_{,\vartheta}\nu_{,R}\right.
\]
\[
\left.-\mu_{R,\vartheta}(\omega\omega_{,R}+\omega^2\psi_{,R})-\mu_{\vartheta,R}(\omega\omega_{,\vartheta}+\omega^2\psi_{,\vartheta})\right]e^{2\psi}
+\frac{1}{4}\omega^2\omega_{,R}\omega_{,\vartheta} e^{4\psi-2\nu},
\]
\[
R_{tR\varphi
\vartheta}=\left[-\frac{1}{2}\omega_{,R,\vartheta}-\omega_{,R}\psi_{,\vartheta}-\frac{1}{2}\omega_{,\vartheta}\psi_{,R}-\omega\psi_{,R,\vartheta}-
\omega\psi_{,R}\psi_{,\vartheta}+\frac{1}{2}\omega_{,R}\nu_{,\vartheta}
\right.
\]
\[
\left.+\mu_{R,\vartheta}\left(\frac{1}{2}\omega_{,R}+\omega\psi_{,R}\right)+\mu_{\vartheta,R}\left(\frac{1}{2}\omega_{,\vartheta}+\omega\psi_{,\vartheta}\right)\right]e^{2\psi}-
\frac{1}{4}\omega\omega_{,R}\omega_{,\vartheta}e^{4\psi-2\nu},
\]
\[
R_{\varphi
Rt\vartheta}=\left[-\frac{1}{2}\omega_{,R,\vartheta}-\frac{1}{2}\omega_{,R}\psi_{,\vartheta}-\omega_{,\vartheta}\psi_{,R}-\omega\psi_{,R,\vartheta}-
\omega\psi_{,R}\psi_{,\vartheta}+\frac{1}{2}\omega_{,\vartheta}\nu_{,R}
\right.
\]
\[
\left.+\mu_{R,\vartheta}\left(\frac{1}{2}\omega_{,R}+\omega\psi_{,R}\right)+\mu_{\vartheta,R}\left(\frac{1}{2}\omega_{,\vartheta}+\omega\psi_{,\vartheta}\right)\right]e^{2\psi}-
\frac{1}{4}\omega\omega_{,R}\omega_{,\vartheta}e^{4\psi-2\nu},
\]
\[
R_{\varphi R\varphi
\vartheta}=\left[e^{-\psi-\mu_\vartheta}\left(e^{\psi-\mu_R}\psi_{,R}\right)_{,\vartheta}-\psi_{,\vartheta}\mu_{\vartheta,R}e^{-\mu_R-\mu_\vartheta}+
\frac{1}{4}\omega_{,R}\omega_{,\vartheta}e^{2\psi-2\nu-\mu_R-\mu_\vartheta}\right]e^{2\psi+\mu_R+\mu_\vartheta},
\]
\[
R_{\vartheta R \vartheta
R}=\left[e^{-\mu_R-\mu_\vartheta}\left(e^{\mu_R-\mu_\vartheta}\mu_{R,\vartheta}\right)_{,\vartheta}+
e^{-\mu_R-\mu_\vartheta}\left(e^{\mu_\vartheta-\mu_R}\mu_{\vartheta,R}\right)_{,R}
\right]e^{2\mu_R+2\mu_\vartheta}.
\]
\subsection{Covariant Riemann tensor components in a non coordinate
basis}\label{vierbeincomp} Starting with the metric (\ref{sam}) we
can take the basis one-forms \cite{chandra1,chandra}
\[
e^{(t)}=e^\nu dt,\quad e^{(\varphi)}=-\omega
e^{\psi}dt+e^{\psi}d\varphi,\quad e^{(R)}=e^{\mu_R}dR,\quad
e^{(\vartheta)}=e^{\mu_\vartheta}d\vartheta,
\]
where we enclosed tetrad indices in a parenthesis in order to
distinguish them from tensor indices. The vierbein associated to
the above choice of the basis one-forms is
\[
e_{(t)}=e^{-\nu}\partial_t+\omega e^{-\nu}\partial_{\varphi},\quad
e_{(\varphi)}=e^{-\psi}\partial_\varphi,\quad
e_{(2)}=e^{-\mu_2}\partial_2,\quad e_{(3)}=e^{-\mu_3}\partial_3.
\]
Moreover, the chosen tetrad satisfies the relation
\[
e_{(a)}{}^{\mu}e_{(b)}{}^{\nu}g_{\mu\nu}=\eta_{(a)(b)},\quad
\eta_{(a)(b)}=\rm{diag}(1,-1,-1,-1).
\]
The tetrad components of the covariant Riemann tensor are given by
\[
R_{(a)(b)(c)(d)}=R_{\mu\nu\rho\lambda}e_{(a)}{}^\mu e_{(b)}{}^\nu
e_{(c)}{}^\rho e_{(d)}{}^\lambda.
\]
The non-vanishing components of the covariant Riemann tensor are
\[
R_{(\varphi)(t)(\varphi)(t)}=\left[e_{(t)}{}^t
e_{(\varphi)}{}^\varphi\right]^2 R_{\varphi t \varphi t}
\]
\[
=-e^{-2\mu_R}\nu_{,R}\psi_{,R}-e^{-2\mu_\vartheta}\nu_{,\vartheta}\psi_{,\vartheta}-\frac{1}{4}e^{2(\psi-\nu)}\left[
e^{-2\mu_R}(\omega_{,R})^2+e^{-2\mu_\vartheta}(\omega_{,\vartheta})^2\right],
\]
\[
R_{(R)(t)(R)(t)}=\left[e_{(t)}{}^t e_{(R)}{}^R\right]^2
R_{RtRt}+2\left[e_{(R)}{}^R\right]^2 e_{(t)}{}^t e_{(t)}{}^\varphi
R_{RtR\varphi}+\left[e_{(t)}{}^\varphi e_{(R)}{}^R\right]^2
R_{R\varphi R\varphi}
\]
\[
=-e^{-\mu_R-\nu}\left(e^{\nu-\mu_R}\nu_{,R}\right)_{,R}-\nu_{,\vartheta}\mu_{R,\vartheta}e^{-2\mu_\vartheta}
+\frac{3}{4}(\omega_{,R})^2e^{2\psi-2\mu_R-2\nu},
\]
\[
R_{(\vartheta)(t)(\vartheta)(t)}=\left[e_{(t)}{}^t
e_{(\vartheta)}{}^\vartheta\right]^2 R_{\vartheta t\vartheta
t}+2\left[e_{(\vartheta)}{}^\vartheta\right]^2 e_{(t)}{}^t
e_{(t)}{}^\varphi R_{\vartheta
t\vartheta\varphi}+\left[e_{(t)}{}^\varphi
e_{(\vartheta)}{}^\vartheta\right]^2
R_{\vartheta\varphi\vartheta\varphi}
\]
\[
=-e^{-\mu_\vartheta-\nu}\left(e^{\nu-\mu_\vartheta}\nu_{,\vartheta}\right)_{,\vartheta}-\nu_{,R}\mu_{\vartheta,R}e^{-2\mu_R}
+\frac{3}{4}(\omega_{,\vartheta})^2e^{2\psi-2\mu_\vartheta-2\nu},
\]
\[
R_{(R)(t)(R)(\varphi)}=\left[e_{(R)}{}^R\right]^2 e_{(t)}{}^t
e_{(\varphi)}{}^\varphi R_{RtR\varphi}+\left[e_{(R)}{}^R\right]^2
e_{(t)}{}^\varphi e_{(\varphi)}{}^\varphi R_{R\varphi R\varphi}
\]
\[
=-\omega_{,R}\left(\psi_{,R}-\frac{1}{2}\nu_{,R}\right)e^{\psi-\nu-2\mu_R}-\frac{1}{2}e^{-\mu_R-\nu}\left(e^{\psi-\mu_R}\omega_{,R}\right)_{,R}-\frac{1}{2}\omega_{,\vartheta}\mu_{R,\vartheta}e^{\psi-\nu-2\mu_\vartheta},
\]
\[
R_{(\vartheta)(t)(\vartheta)(\varphi)}=\left[e_{(\vartheta)}{}^\vartheta\right]^2
e_{(t)}{}^t e_{(\varphi)}{}^\varphi R_{\vartheta
t\vartheta\varphi}+\left[e_{(\vartheta)}{}^\vartheta\right]^2
e_{(t)}{}^\varphi e_{(\varphi)}{}^\varphi R_{\vartheta\varphi
\vartheta\varphi}
\]
\[
=-\omega_{,\vartheta}\left(\psi_{,\vartheta}-\frac{1}{2}\nu_{,\vartheta}\right)e^{\psi-\nu-2\mu_\vartheta}-\frac{1}{2}e^{-\mu_\vartheta-\nu}\left(e^{\psi-\mu_\vartheta}\omega_{,\vartheta}\right)_{,\vartheta}-\frac{1}{2}\omega_{,R}\mu_{\vartheta,R}e^{\psi-\nu-2\mu_R},
\]
\[
R_{(R)(\varphi)(R)(\varphi)}=\left[e_{(R)}{}^R
e_{(\varphi)}{}^\varphi\right]^2 R_{R\varphi R\varphi}
\]
\[
=e^{-\psi-\mu_R}\left(e^{\psi-\mu_R}\psi_{,R}\right)_{,R}+\mu_{R,\vartheta}\psi_{,\vartheta}e^{-2\mu_\vartheta}+\frac{1}{4}\omega_{,R}^2
e^{2\psi-2\mu_R-2\nu},
\]
\[
R_{(\vartheta)(\varphi)(\vartheta)(\varphi)}=\left[e_{(\vartheta)}{}^\vartheta
e_{(\varphi)}{}^\varphi\right]^2 R_{\vartheta\varphi
\vartheta\varphi}
\]
\[
=e^{-\psi-\mu_\vartheta}\left(e^{\psi-\mu_\vartheta}\psi_{,\vartheta}\right)_{,\vartheta}+\mu_{\vartheta,R}\psi_{,R}e^{-2\mu_R}+\frac{1}{4}\omega_{,\vartheta}^2
e^{2\psi-2\mu_\vartheta-2\nu},
\]
\[
R_{(t)(R)(t)(\vartheta)}=\left[e_{(t)}{}^t\right]^2 e_{(R)}{}^R
e_{(\vartheta)}{}^\vartheta R_{tRt\vartheta}+(R_{tR\varphi
\vartheta}+R_{\varphi Rt\vartheta})e_{(t)}{}^t e_{(t)}{}^\varphi
e_{(R)}{}^R e_{(\vartheta)}{}^\vartheta
\]
\[
+\left[e_{(t)}{}^\varphi\right]^2 e_{(R)}{}^R
e_{(\vartheta)}{}^\vartheta R_{\varphi R\varphi\vartheta}
\]
\[
=-e^{-\nu-\mu_\vartheta}\left(e^{\nu-\mu_R}\nu_{,R}\right)_{,\vartheta}+\nu_{,\vartheta}\mu_{\vartheta,R}
e^{-\mu_R-\mu_\vartheta}+\frac{3}{4}\omega_{,R}\omega_{,\vartheta}e^{2\psi-2\nu-\mu_{R}-\mu_{\vartheta}},
\]
\[
R_{(\varphi)(R)(\varphi)(\vartheta)}=\left[e_{(\varphi)}{}^\varphi\right]^2
e_{(R)}{}^R e_{(\vartheta)}{}^\vartheta R_{\varphi
R\varphi\vartheta}
\]
\[
=e^{-\psi-\mu_\vartheta}\left(e^{\psi-\mu_R}\psi_{,R}\right)_{,\vartheta}-\psi_{,\vartheta}\mu_{\vartheta,R}e^{-\mu_R-\mu_\vartheta}+
\frac{1}{4}\omega_{,R}\omega_{,\vartheta}e^{2\psi-2\nu-\mu_R-\mu_\vartheta},
\]
\[
R_{(\vartheta)(R)(\vartheta)(R)}=\left[e_{(R)}{}^R
e_{(\vartheta)}{}^\vartheta \right]^2 R_{\vartheta R\vartheta R}=
e^{-\mu_R-\mu_\vartheta}\left[\left(e^{\mu_R-\mu_\vartheta}\mu_{R,\vartheta}\right)_{,\vartheta}+
\left(e^{\mu_\vartheta-\mu_R}\mu_{\vartheta,R}\right)_{,R}\right].
\]
\subsection{Ricci tensor components}
The tetrad components of the Ricci tensor have been computed
according to the formula
\[
R_{(a)(b)}=\eta^{(c)(d)}R_{(c)(a)(d)(b)}.
\]
The non-vanishing components are
\[
R_{(t)(t)}=-R_{(\varphi)(t)(\varphi)(t)}-R_{(R)(t)(R)(t)}-R_{(\vartheta)(t)(\vartheta)(t)}
\]
\[
=e^{-2\mu_R}\left[\nu_{,R,R}+\nu_{,R}(\psi+\nu-\mu_R+\mu_\vartheta)_{,R}\right]
+e^{-2\mu_\vartheta}\left[\nu_{,\vartheta,\vartheta}+\nu_{,\vartheta}(\psi+\nu+\mu_R-\mu_\vartheta)_{,\vartheta}\right]
\]
\begin{equation}\label{RTT}
-\frac{1}{2}e^{2(\psi-\nu)}\left[(\omega_{,R})^2e^{-2\mu_R}+(\omega_{,\vartheta})^2e^{-2\mu_\vartheta}\right],
\end{equation}
\[
R_{(t)(\varphi)}=-R_{(R)(t)(R)(\varphi)}-R_{(\vartheta)(t)(\vartheta)(\varphi)}
\]
\begin{equation}\label{RTF}
=\frac{1}{2}e^{-2\psi-\mu_R-\mu_\vartheta}\left[\left(\omega_{,R}e^{3\psi-\nu-\mu_R+\mu_\vartheta}\right)_{,R}+
\left(\omega_{,\vartheta}e^{3\psi-\nu-\mu_\vartheta+\mu_R}\right)_{,\vartheta}\right],
\end{equation}
\[
R_{(\varphi)(\varphi)}=R_{(\varphi)(t)(\varphi)(t)}-R_{(R)(\varphi)(R)(\varphi)}-R_{(\vartheta)(\varphi)(\vartheta)(\varphi)}
\]
\[
=-e^{-2\mu_R}\left[\psi_{,R,R}+\psi_{,R}(\psi+\nu-\mu_R+\mu_\vartheta)_{,R}\right]-
e^{-2\mu_\vartheta}\left[\psi_{,\vartheta,\vartheta}+\psi_{,\vartheta}(\psi+\nu+\mu_R-\mu_\vartheta)_{,\vartheta}\right]
\]
\begin{equation}\label{RFF}
-\frac{1}{2}e^{2(\psi-\nu)}\left[(\omega_{,R})^2e^{-2\mu_R}+(\omega_{,\vartheta})^2e^{-2\mu_\vartheta}\right],
\end{equation}
\[
R_{(R)(\vartheta)}=R_{(t)(R)(t)(\vartheta)}-R_{(\varphi)(R)(\varphi)(\vartheta)}
=-e^{-\mu_R-\mu_\vartheta}\cdot
\]
\begin{equation}\label{R23}
\left[(\psi+\nu)_{,R,\vartheta}-(\psi+\nu)_{,R}\mu_{R,\vartheta}-(\psi+\nu)_{,\vartheta}\mu_{\vartheta,R}+
\psi_{,R}\psi_{,\vartheta}
+\nu_{,R}\nu_{,\vartheta}-\frac{1}{2}\omega_{,R}\omega_{,\vartheta}e^{2(\psi-\nu)}\right],
\end{equation}
\[
R_{(R)(R)}=-R_{(t)(R)(t)(R)}-R_{(\varphi)(R)(\varphi)(R)}-R_{(\vartheta)(R)(\vartheta)(R)}
\]
\[
=-e^{-\mu_R-\nu}\left(e^{\nu-\mu_R}\nu_{,R}\right)_{,R}-e^{-\mu_R-\psi}\left(e^{\psi-\mu_R}\psi_{,R}\right)_{,R}
-e^{-2\mu_\vartheta}(\psi+\nu)_{,\vartheta}\mu_{R,\vartheta}
\]
\[
-e^{-\mu_R-\mu_\vartheta}\left[\left(e^{\mu_R-\mu_\vartheta}\mu_{R,\vartheta}\right)_{,\vartheta}+
\left(e^{\mu_\vartheta-\mu_R}\mu_{\vartheta,R}\right)_{,R}\right]+\frac{1}{2}\left(\omega_{,R}\right)^2
e^{2\psi-2\mu_R-2\nu},
\]
\[
R_{(\vartheta)(\vartheta)}=-R_{(t)(\vartheta)(t)(\vartheta)}-R_{(\varphi)(\vartheta)(\varphi)(\vartheta)}
-R_{(R)(\vartheta)(R)(\vartheta)}
\]
\[
=-e^{-\mu_\vartheta-\nu}\left(e^{\nu-\mu_\vartheta}\nu_{,\vartheta}\right)_{,\vartheta}-e^{-\mu_\vartheta-\psi}
\left(e^{\psi-\mu_\vartheta}\psi_{,\vartheta}\right)_{,\vartheta}-e^{-2\mu_R}(\psi+\nu)_{,R}\mu_{\vartheta,R}
\]
\[
-e^{-\mu_R-\mu_\vartheta}\left[\left(e^{\mu_R-\mu_\vartheta}\mu_{R,\vartheta}\right)_{,\vartheta}+
\left(e^{\mu_\vartheta-\mu_R}\mu_{\vartheta,R}\right)_{,R}\right]+\frac{1}{2}\left(\omega_{,\vartheta}\right)^2
e^{2\psi-2\mu_\vartheta-2\nu}.
\]
Finally, the components $G_{(R)(R)}$ and
$G_{(\vartheta)(\vartheta)}$ of the Einstein tensor are computed
to be
\[
G_{(R)(R)}=e^{-2\mu_R}\left[\nu_{,R}\left(\psi+\mu_\vartheta\right)_{,R}+\psi_{,R}\mu_{\vartheta,R}\right]
+\frac{1}{4}e^{2\psi-2\nu}\left[(\omega_{,R})^2
e^{-2\mu_R}-(\omega_{,\vartheta})^2 e^{-2\mu_\vartheta}\right]
\]
\begin{equation}\label{GRR}
+e^{-2\mu_\vartheta}\left[\left(\psi+\nu\right)_{,\vartheta,\vartheta}+(\psi+\nu)_{,\vartheta}(\nu-\mu_\vartheta)_{,\vartheta}
+\psi_{,\vartheta}\psi_{,\vartheta}\right],
\end{equation}
\[
G_{(\vartheta)(\vartheta)}=e^{-2\mu_R}\left[\left(\psi+\nu\right)_{,R,R}+(\psi+\nu)_{,R}(\nu-\mu_R)_{,R}
+\psi_{,R}\psi_{,R}\right]
\]
\begin{equation}\label{GTHTH}
+e^{-2\mu_\vartheta}\left[\nu_{,\vartheta}\left(\psi+\mu_R\right)_{,\vartheta}+\psi_{,\vartheta}\mu_{R,\vartheta}\right]
-\frac{1}{4}e^{2\psi-2\nu}\left[(\omega_{,R})^2
e^{-2\mu_R}-(\omega_{,\vartheta})^2 e^{-2\mu_\vartheta}\right].
\end{equation}
\section{Kerrr metric revisited}
We show that the so-called Kerrr metric \cite{spallucci2}
\[
ds^2=\left(1-\frac{2RM(R)}{\Sigma}\right)dt^2+\frac{4aRM(R)}{\Sigma}\sin^2{\vartheta}dtd\varphi-\frac{\Sigma}{\Delta}dR^2
-\Sigma d\vartheta^2
\]
\begin{equation}\label{metrica1}
-\frac{\sin^2{\vartheta}}{\Sigma}\left[(R^2+a^2)^2-a^2\sin^2{\vartheta}\Delta\right]d\varphi^2
\end{equation}
with
\[
\Sigma=R^2+a^2\cos^2{\vartheta},\quad\Delta=R^2-2RM(R)+a^2
\]
does not satisfy Einstein field equations $G_{\mu\nu}=-8\pi
T_{\mu\nu}$ with energy-momentum tensor
\begin{equation}\label{tensore en mom}
T^{\mu}{}_{\nu}=(\rho+p_\vartheta)(u^\mu
u_\nu-\ell^\mu\ell_\nu)-p_\vartheta\delta^\mu_\nu,
\end{equation}
\begin{equation}\label{componenti}
u^\mu=\sqrt{-g^{RR}}\left(\delta_t^\mu+\Omega\delta^\varphi_\mu\right),\quad
\frac{u^\varphi}{u^t}=\Omega=\frac{a}{R^2+a^2},\quad
\ell^\mu=-\frac{1}{\sqrt{-g_{RR}}}\delta_{R}^\mu
\end{equation}
and $\rho$ given by
\begin{equation}\label{densita}
\rho(R,\vartheta)=\frac{R^4}{\Sigma^2}\rho_G(R),\quad
\rho_G(R)=\frac{M}{8\pi^{3/2}\ell_0^3}e^{-R^2/4\ell^2_0},\quad\rho+p_\vartheta=-\frac{\Sigma}{2R}\partial_R\rho.
\end{equation}
In contrast to the theorems 1 and 2 we use here the
four-velocities as given in \cite{spallucci2}. Notice that
$\rho_G$ is a function depending uniquely on the spatial variable
$R$. For the present purpose it is sufficient to consider the
Einstein field equations
\[
G_{tt}=-8\pi T_{tt},\quad G_{RR}=-8\pi T_{RR}.
\]
By means of the software package Maple 12 we found that the
components $G_{tt}$ and $G_{RR}$ of the Einstein tensor are given
by
\begin{equation}\label{GTT}
G_{tt}=\frac{1}{\Sigma^3}\left[a^2
R\Sigma\sin^2{\vartheta}\frac{d^2
M}{dR^2}-2(R^2\Delta-a^4\sin^2{\vartheta}\cos^2{\vartheta})\frac{dM}{dR}\right],
\end{equation}
\begin{equation}\label{comp_RR}
G_{RR}=\frac{2R^2}{\Sigma\Delta}\frac{dM}{dR}.
\end{equation}
Making use of (\ref{tensore en mom}), (\ref{componenti}) and
(\ref{densita}) we find that
\[
T_{tt}=-\frac{\Delta^3}{2R(R^2+a^2)^2}\partial_R\left(\frac{R^4}{\Sigma^2}\rho_G(R)\right)
\]
\[
+\left(1-\frac{2RM(R)}{\Sigma}\right)\left[\frac{R^4}{\Sigma^2}\rho_G(R)+\frac{\Sigma}{2R}\partial_R\left(\frac{R^4}{\Sigma^2}\rho_G(R)\right)\right],\quad
T_{RR}=-\frac{R^4}{\Sigma\Delta}\rho_G(R).
\]
As in \cite{spallucci2} from the equation $G_{RR}=-8\pi T_{RR}$ we
obtain the following result for the mass function $M(R)$, namely
\begin{equation}\label{mass function}
M(r)=4\pi\int_0^R dx~x^2\rho_G(x).
\end{equation}
Concerning the equation $G_{tt}=-8\pi T_{tt}$ it is convenient to
rewrite the component $T_{tt}$ as follows
\[
T_{tt}=-\frac{\Delta^3}{2R(R^2+a^2)^2}\left(4\frac{R^2}{\Sigma^2}\rho_G(R)-4\frac{R^4}{\Sigma^3}\rho_G(R)+
\frac{R^3}{\Sigma^2}\frac{d\rho_G}{dR} \right)+
\]
\begin{equation}\label{TTT}
\left(1-\frac{2RM(R)}{\Sigma}\right)\left(2\frac{R^2}{\Sigma}\rho_G(R)-\frac{R^4}{\Sigma^2}\rho_G(R)+\frac{R^3}{2\Sigma}\frac{d\rho_G}{dR}
\right).
\end{equation}
Equating (\ref{GTT}) to $-8\pi$(\ref{TTT}) and multiplying by
$\Sigma^3$ we obtain
\[
a^2 R\Sigma\sin^2{\vartheta}\frac{d^2
M}{dR^2}-2(R^2\Delta-a^4\sin^2{\vartheta}\cos^2{\vartheta})\frac{dM}{dR}
\]
\[
=8\pi
R^2\left[\frac{2\Delta^3(\Sigma-R^2)}{(R^2+a^2)^2}-(\Sigma-2RM(R))(2\Sigma-R^2)\right]\rho_G(R)+
\]
\[
4\pi
R^3\Sigma\left[\frac{\Delta^3}{(R^2+a^2)^2}-(\Sigma-2RM(R))\right]\frac{d\rho_G}{dR}.
\]
In order to further simplify the above expression we shall apply
(\ref{mass function}) to the l.h.s. of the above equation.
Therefore, taking into account that
\[
a^2 R\Sigma\sin^2{\vartheta}\frac{d^2
M}{dR^2}-2(R^2\Delta-a^4\sin^2{\vartheta}\cos^2{\vartheta})\frac{dM}{dR}
\]
\[
=8\pi
R^2\left[2a^4\sin^2{\vartheta}\cos^2{\vartheta}-R^2(\Sigma-2RM(R))\right]\rho_G(R)+
4\pi a^2 R^3 \Sigma\sin^2{\vartheta}\frac{d\rho_G}{dR},
\]
we finally obtain the equation
\[
8\pi
R^2\left[\frac{2\Delta^3(\Sigma-R^2)}{(R^2+a^2)^2}-(\Sigma-2RM(R))(2\Sigma-R^2)+R^2(\Sigma-2RM(R))\right.
\]
\[
\left.-2a^4\sin^2{\vartheta}\cos^2{\vartheta}\right]\rho_G(R)+4\pi
R^2\left[\frac{\Delta^3\Sigma}{(R^2+a^2)^2}\right.
\]
\begin{equation}\label{penultima}
\left.-\Sigma(\Sigma-2RM(R))-a^2\Sigma\sin^2{\vartheta}\right]\frac{d\rho_G}{dR}=0.
\end{equation}
Observing that
\[
-(\Sigma-2RM(R))(2\Sigma-R^2)+R^2(\Sigma-2RM(R))-2a^4\sin^2{\vartheta}\cos^2{\vartheta}=
-2a^2\Delta\cos^2{\vartheta}
\]
and
\[
-(\Sigma-2RM(R))(2\Sigma-R^2)-a^2\Sigma\sin^2{\vartheta}=-\Sigma\Delta
\]
equation (\ref{penultima}) simplifies as follows
\begin{equation}\label{ultima}
4\left[\frac{\Delta^2(\Sigma-R^2)}{(R^2+a^2)^2}-a^2\cos^2{\vartheta}\right]\rho_G(R)+
R\Sigma\left[\frac{\Delta^2}{(R^2+a^2)^2}-1\right]\frac{d\rho_G}{dR}=0.
\end{equation}
Since
\[
\frac{\Delta^2(\Sigma-R^2)}{(R^2+a^2)^2}-a^2\cos^2{\vartheta}=a^2\cos^2{\vartheta}\left[\frac{\Delta^2}{(R^2+a^2)^2}-1\right]
\]
equation (\ref{ultima}) takes the final form
\[
\frac{d\rho_G}{dR}+\frac{4a^2\cos^2{\vartheta}}{R\Sigma}\rho_G(R)=0.
\]
Hence, we have a contradiction since it has been assumed that
$\rho_G$ depends uniquely on the variable $R$.

\section{Choices of the energy densities}
Throughout the paper we have kept the density $\rho_\theta$ as a
free function. However, a few comments regarding the choice are in
order. The authors of (\cite{spallucci2}) advocate the following
form
\[
\rho_\theta(R,\vartheta)=\frac{M}{8\pi^{3/2}\ell_0^{3/2}}\frac{R^2}{\Sigma}~e^{-R^2/4\ell^2_0},
\]
where $\Sigma=R^2+a^2\cos^{2}{\vartheta}$, $a$ is the angular
momentum of the black hole per unit mass and $R,\vartheta,\varphi$
are Boyer-Lindquist coordinates. According to the prescription
outlined in the introduction it appears to us that a more
sophisticated choice might also be possible. Consider the
classical ring singularity
\[
\rho_{cl}(x,y,z)=M\delta(x^2+y^2-a^2)\delta(z).
\]
Then, every $\delta$-distribution must be replaced by a separate
Gaussian distribution. This would give us in cylindrical
coordinates a density proportional to
\[
\rho(\xi,z)\propto e^{-(\xi-a)^2/4\theta}e^{-z^2/4\theta},
\]
where $\xi=\sqrt{x^2+y^2}$. Of course, this equation is written in
cylindrical coordinates and should be rewritten in Boyer-Lindquist
coordinates.


\newpage
\begin{thebibliography}{999}
\bibitem{Rest1}
Madore J 2000 \textit{An introduction to noncommutative geometry}
in ``Geometry and Quantum Physics'', Lectures Notes in Physics
Vol. 543, Springer-Verlag; Chamseddine A H, Felder G and
Fr\"{o}hlich J 2005 \textit{Gravity in non-commutative geometry'}
Commun. Math. Phys. {\bf{155}}, 1993; Connes A 1995
\textit{Noncommutative geometry and reality}, J. Math. Phys. {\bf
36}, 6194

\bibitem{three}
Alvarez Gaume L and Wadia S R 2001 \textit{Gauge theory on a
quantum phase space} Phys. Lett. B {\bf{501}}, 319; Alvarez Gaume
L and Barbon J L F 2001 \textit{Nonlinear vacuum phenomena in
noncommutative QED} Int. J. Mod. Phys. A {\bf{16}}, 1123

\bibitem{four}
Weyl H 1927 \textit{Quantenmechanik und Gruppentheorie} Z. Phys.
{\bf{46}} 1; Wigner E P 1932 \textit{On the Quantum Correction For
Thermodynamic Equilibrium} Phys. Rev. {\bf{40}}, 749; Moyal G E
1949 \textit{Quantum mechanics as a statistical theory} Proc.
Camb. Phil. Soc. {\bf{45}},  99

\bibitem{peet}
Peet A W and Polchinski J 1999 \textit{UV-IR relations in AdS
dynamics} Phys. Rev. D {\bf{59}}, 065011

\bibitem{spallucci1}
Smailagic A and Spallucci E 2003 \textit{UV divergence-free QFT on
noncommutative plane} J. Phys. A: Math. Gen. {\bf{36}}, 517

\bibitem{Glauber}
Glauber R J 1963 \textit{Coherent and Incoherent States of the
Radiation Field} Phys. Rev. {\bf{131}}, 2766

\bibitem{snyder}
Snyder H S 1949 \textit{Quantized Space-Time} Phys. Rev.
{\bf{71}}, 38

\bibitem{spallucci11}
Smailagic A and Spallucci E 2003 \textit{Feynman Path Integral on
the Noncommutative Plane } J. Phys. A: Math. Gen. {\bf{36}}, 467

\bibitem{indian}
Banerjee R, Majhi B R and Samanta S 2008 \textit{Noncommutative
Black Hole Thermodynamics} Phys. Rev. D {\bf{77}}, 124035;
Banerjee R, Majhi B R and Modak S K 2009 \textit{Noncommutative
Schwarzschild Black Hole and Area Law} Class. Quant. Grav. {\bf{
26}}, 085010; Banerjee R, Gangopadhyay S and Modak S K 2010
\textit{Voros product, Noncommutative Schwarzschild Black Hole and
Corrected Area Law}, Phys. Lett. B {\bf 686}, 181


\bibitem{nico1}
Nicolini P and Spallucci E 2010 \textit{Noncommutative geometry
inspired dirty black holes} Class. Quant. Grav. {\bf{27}}, 015010

\bibitem{nico2}
Nicolini P 2009 \textit{Noncommutative Black Holes, The Final
Appeal To Quantum Gravity: A Review} Int. J. Mod. Phys. A
{\bf{24}}, 1229

\bibitem{spallucci3}
Spallucci E, Smailagic A and Nicolini P 2009
\textit{Non-commutative geometry inspired higher-dimensional
charged, black holes} Phys. Lett. B {\bf{670}}, 449


\bibitem{ansoldi1}
Ansoldi S, Nicolini P, Smailagic A and Spallucci E 2007
\textit{Noncommutative geometry inspired charged black holes}
Phys. Lett. B {\bf{645}}, 261

\bibitem{nico3}
Nicolini P, Smailagic A and Spallucci E 2006
\textit{Noncommutative geometry inspired Schwarzschild black hole}
Phys. Lett. B {\bf{632}}, 547

\bibitem{uns}
Arraut I, Batic D and Nowakowski M 2010 \textit{Maximal extension
of the Schwarzschild space-time inspired by noncommutative
geometry} J. Math. Phys. {\bf{51}}, 022503; Arraut I, Batic D and
Nowakowski M 2009 \textit{A noncommutative model for a mini black
hole} Class. Quantum Grav. {\bf{26}}, 245006

\bibitem{chandra1}
Chandrasekhar S and Friedman J L 1972 \textit{On the stability of
axisymmetric systems to axisymmetric perturbations in General
Relativity. I} Astrophys. J. {\bf 175}, 379


\bibitem{spallucci2}
Smailagic A and Spallucci E 2010 \textit{"Kerrr" black hole: the
Lord of the String} Phys. Lett. B {\bf 688}, 82

\bibitem{chandra}
Chandrasekhar S 1983 \textit{The Mathematical Theory of Black
Holes}, Clarendon Press

\bibitem{Boyer}
Boyer R H and Lindquist R W 1967 \textit{Maximal Analytic
Extension of the Kerr Metric} J. Math. Phys {\bf 8}, 265

\bibitem{Ple}
Plebanski J and Krasinski A 2006 \textit{An Introduction to
General Relativity and Cosmology}, Cambridge University Press

\end{thebibliography}
\end{document}